\documentclass[useAMS,usenatbib]{mn2e}
\usepackage{graphicx}
\usepackage{epstopdf}
\usepackage{xcolor}
\usepackage[percent]{overpic}
\usepackage{mathptmx}
\usepackage{anyfontsize}
\usepackage{t1enc}
\usepackage{url}
\usepackage{lineno}
\title{Inference of Magnetic Field in the Coronal Streamer Invoking Kink Wave Motions generated by Multiple EUV Waves}
\author[Abhishek K. Srivastava, Talwinder Singh, Leon Ofman, and Bhola N. Dwivedi]{A.K.~Srivastava$^{1}$\thanks{E-mail: asrivastava.app@iitbhu.ac.in}, Talwinder Singh$^{1}$, Leon Ofman$^{2}$, and Bhola N. Dwivedi$^{1}$\\
$^{1}$1 Department of Physics, Indian Institute of Technology (Banaras Hindu University), Varanasi-221005, India
\\
$^{2}$Catholic University of America and NASA Goddard Space Flight Center, Solar Physics Laboratory, Code 671,Greenbelt, MD 20771, USA
}
\begin{document}

\date{26 May 2016}

\pagerange{\pageref{firstpage}--\pageref{lastpage}} \pubyear{2016}

\maketitle

\label{firstpage}

\begin{abstract}
Invoking methodology of MHD seismology by observed kink waves, the magnetic field profile within a coronal streamer has been investigated. 
STEREO-B/EUVI temporal image data on 7 March 2012 shows an evolution of two consecutive extreme ultraviolet (EUV) waves that interact with the footpoint of a coronal streamer evident clearly in the co-spatial and co-temporal STEREO-B/COR-I observations. The evolution of consecutive EUV waves is clearly evident in the STEREO-B/EUVI, and its energy exchange with coronal streamer generates kink oscillations.
We estimate the phase velocities of the kink wave perturbations by tracking it at different heights of the coronal streamer. We also estimate the electron densities inside and outside the streamer using spherically symmetric inversion of polarized brightness images in STEREO-B/COR-1 observations. Taking into account the MHD theory of kink waves in a cylindrical waveguide, their observed properties at various heights, and density contrast of the streamer, we estimate the radial profile of magnetic field within this magnetic structure. Both the kink waves diagnose the exponentially decaying radial profiles of the magnetic field in coronal steamer upto 3 solar radii. Within the limit of uncertainties in the measurements, it is indicated that coronal 
magnetic field of the streamer varies slowly in time at various heights, although its nature always remains exponentially decaying. It is seen that during the evolution of second kink motion in the streamer, it increases in brightness (thus mass density), and also in areal extent slightly, which may be associated with the decreased photospheric magnetic flux at the footpoint of the streamer. As a result, the magnetic field profile produced by the second kink wave is reduced  within the streamer compared to the one diagnosed by the first one. The precisely estimated magnetic field profiles with the uncertainty less than 10\%, using principle of MHD seismology, match reasonably well with the empirical profile as well as various observational estimations of the 
outer coronal magnetic fields.  
\end{abstract}


\section{Introduction}

\begin{figure*}
\caption{EUV wavefront as seen in EUVI 195 \r{A} and COR-1 field of view of STEREO-B observations on 7 March 2012. CDs refer to COR-1 disturbances, while EDs depict EUV disturbances. First and second EUV waves are triggered respectively at 00:05 UT and 01:05 UT. The black circle 
at 1.6 solar radii (bottom-middle snapshot) as shown by arrow is the path along which the kinematics of the EUV wave fronts in COR-1 observations
are estimated.}
\begin{tabular}{c c c}
\begin{overpic}[scale=0.03,angle=90,width=.6\textwidth,height=5.4cm,keepaspectratio]{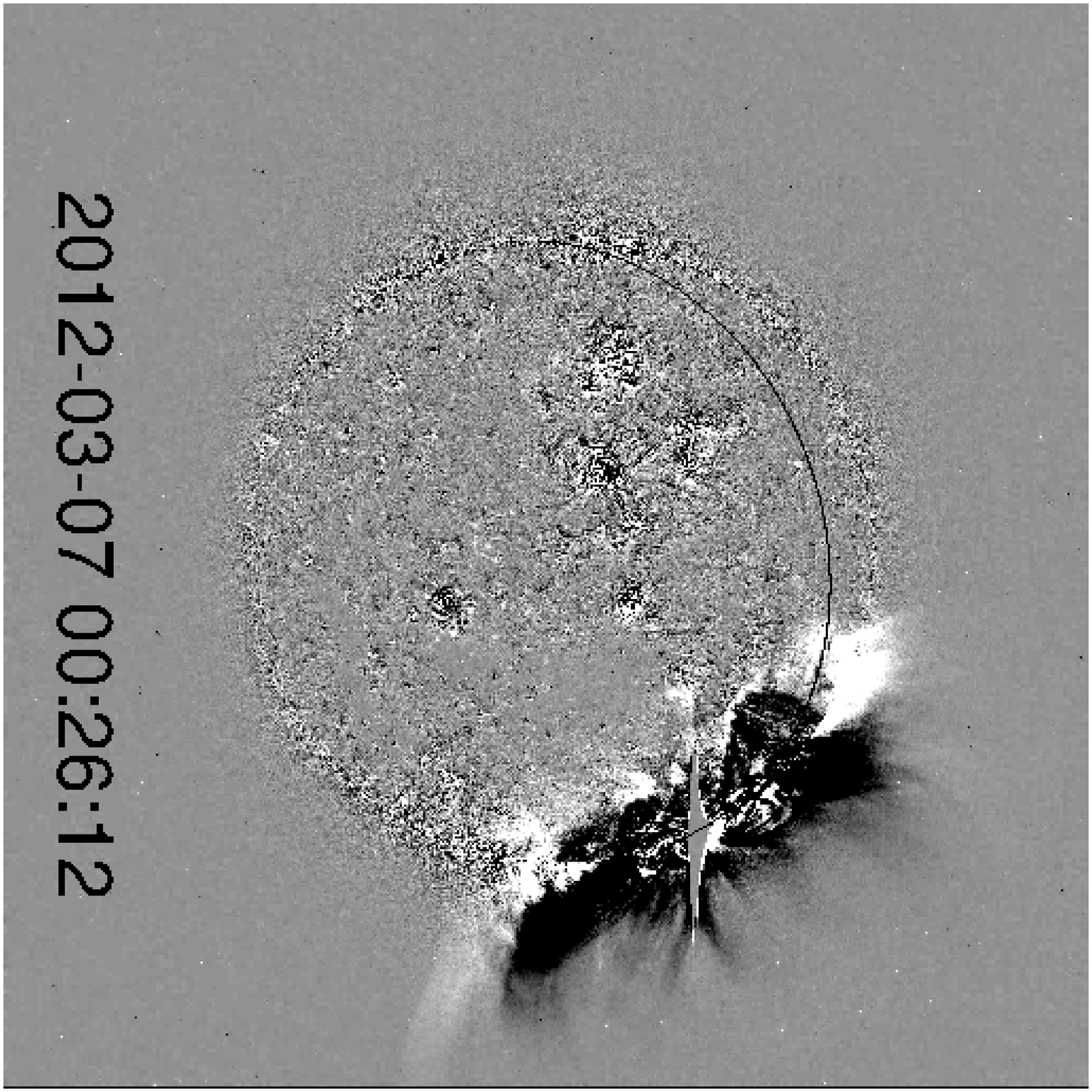}
\put(5.0,88){\color{black}{ \fontsize{4}{5}\selectfont Traced Path 1}}
\put(20,86){\color{black}\vector(1,-1){15}}
\end{overpic}

\includegraphics[scale=0.4,angle=90,width=5.4cm,height=5.4cm,keepaspectratio]{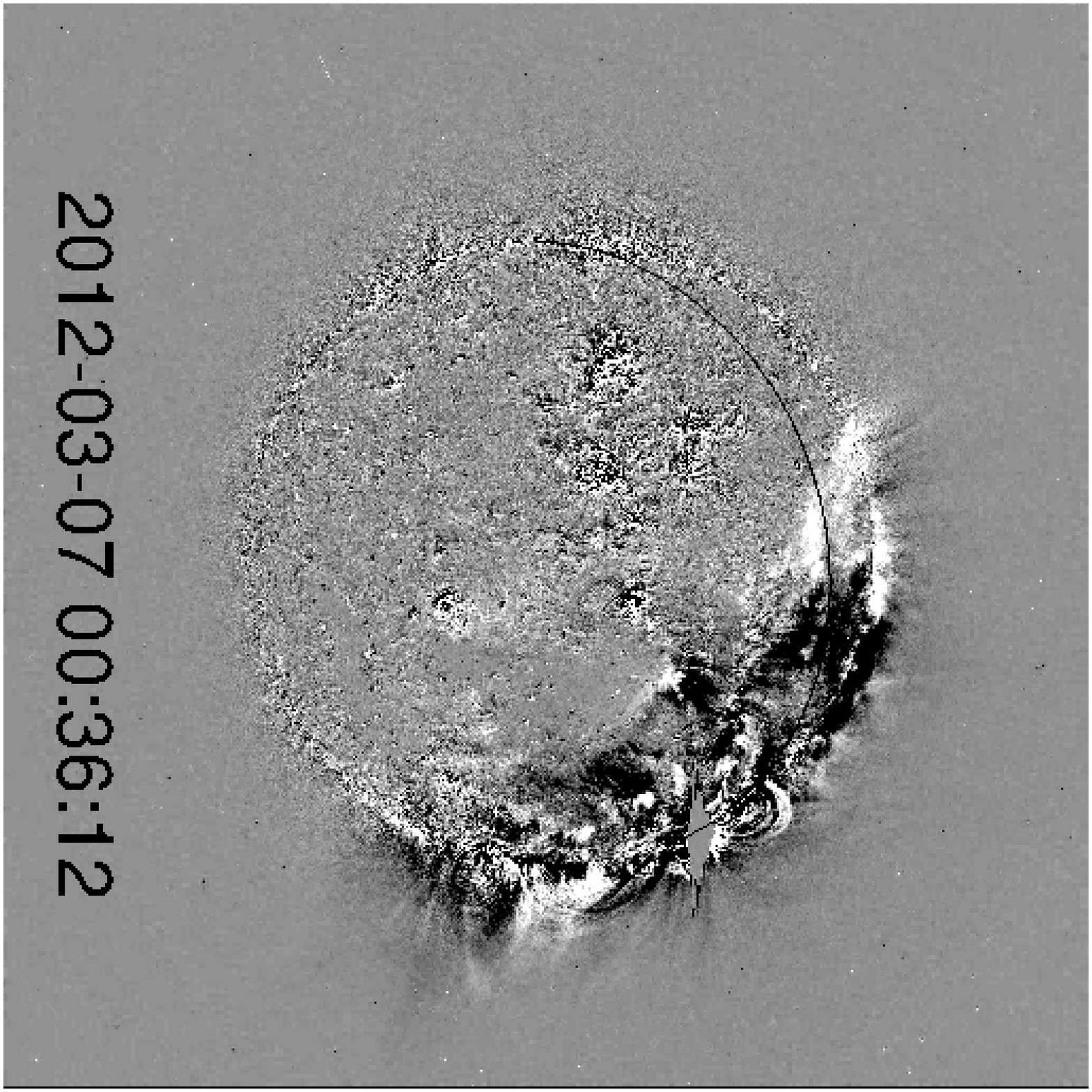}

\includegraphics[scale=0.4,angle=90,width=5.4cm,height=5.4cm,keepaspectratio]{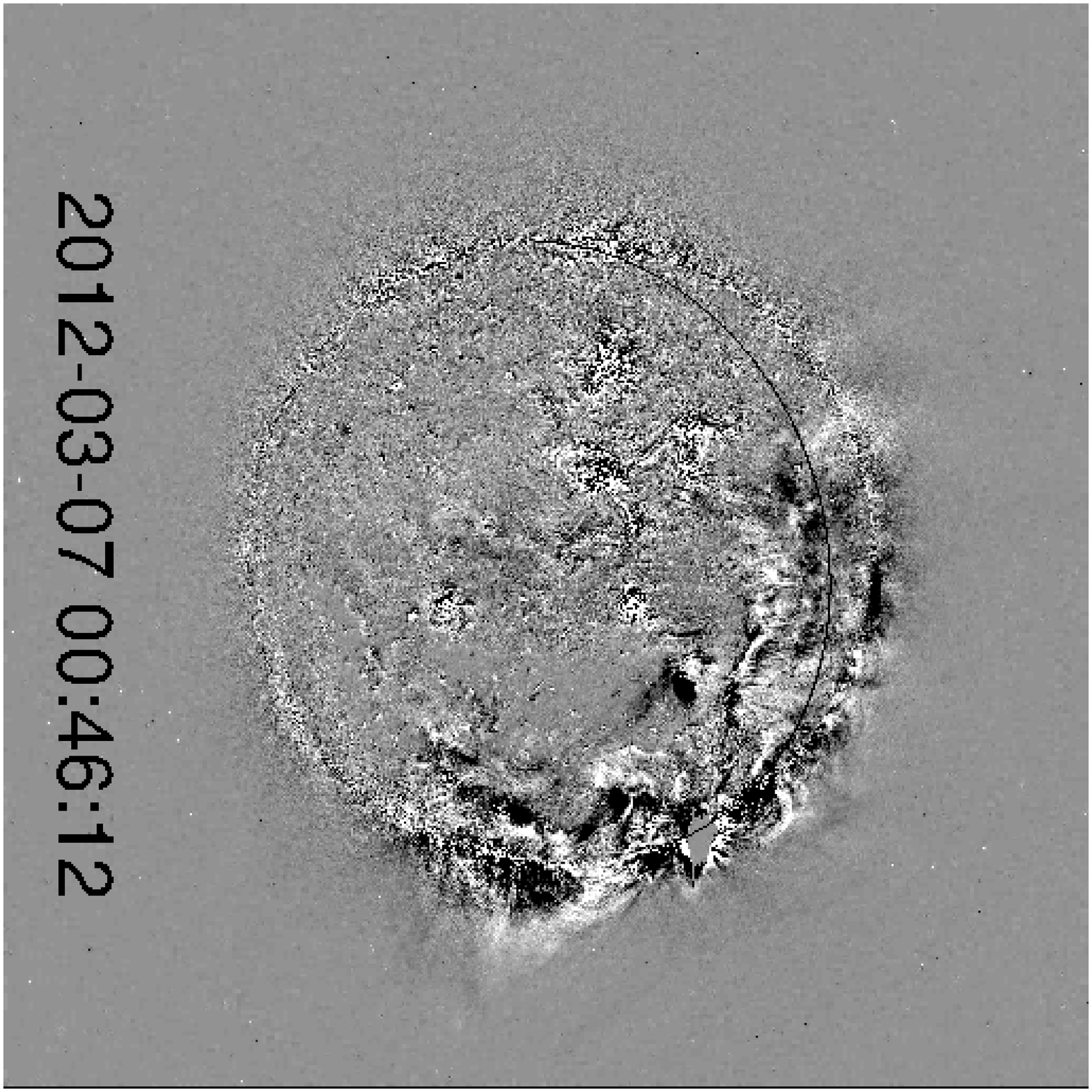} \\
\begin{overpic}[scale=0.03,angle=90,width=.6\textwidth,height=5.4cm,keepaspectratio]{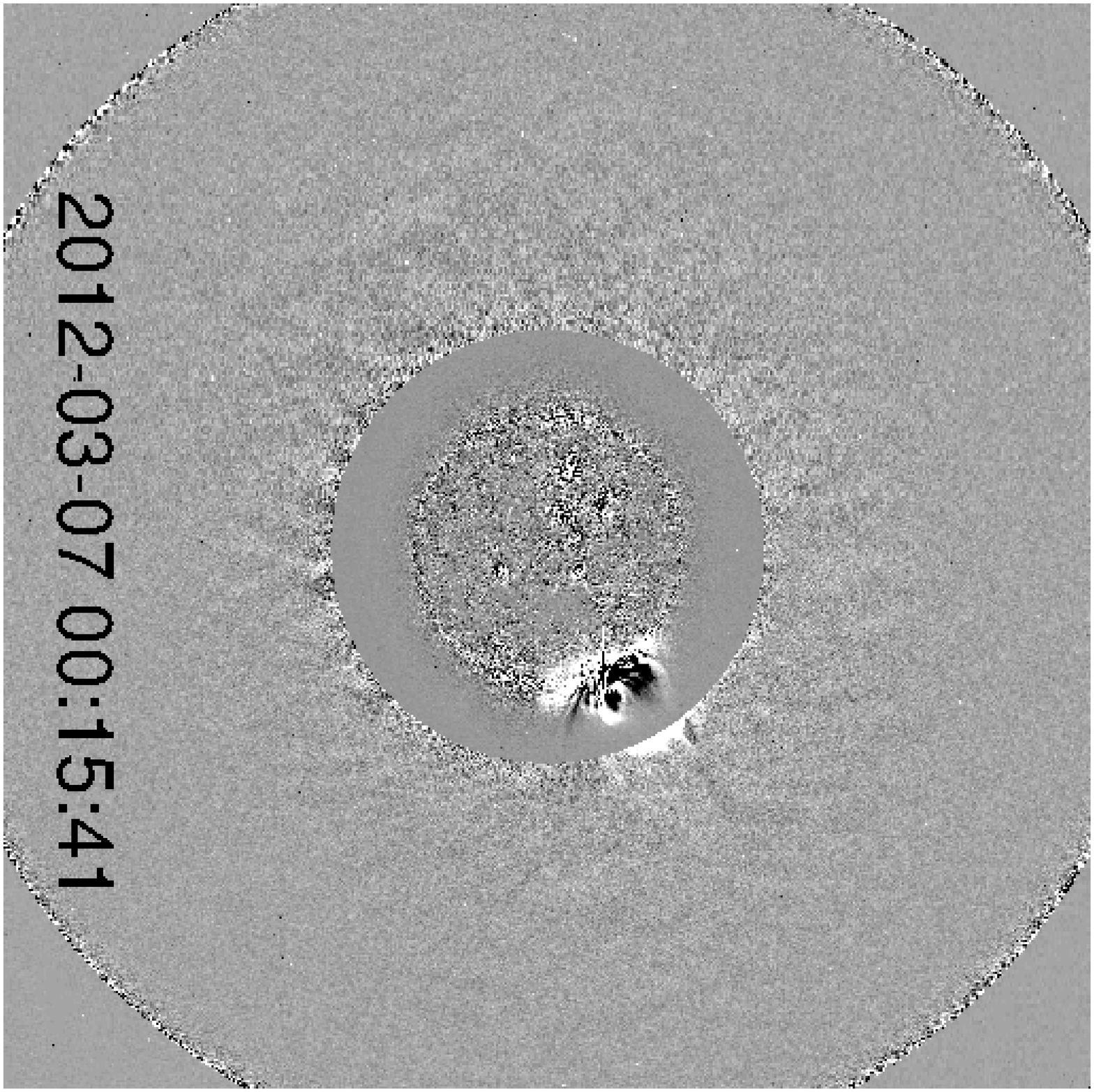}
\put(70,72){\color{black}{ \fontsize{4}{5}\selectfont Flare Source}}
\put(77,70){\color{black}\vector(-1,-1){15}}
\end{overpic}

\begin{overpic}[scale=0.03,angle=90,width=5.4cm, height=5.4cm,keepaspectratio]{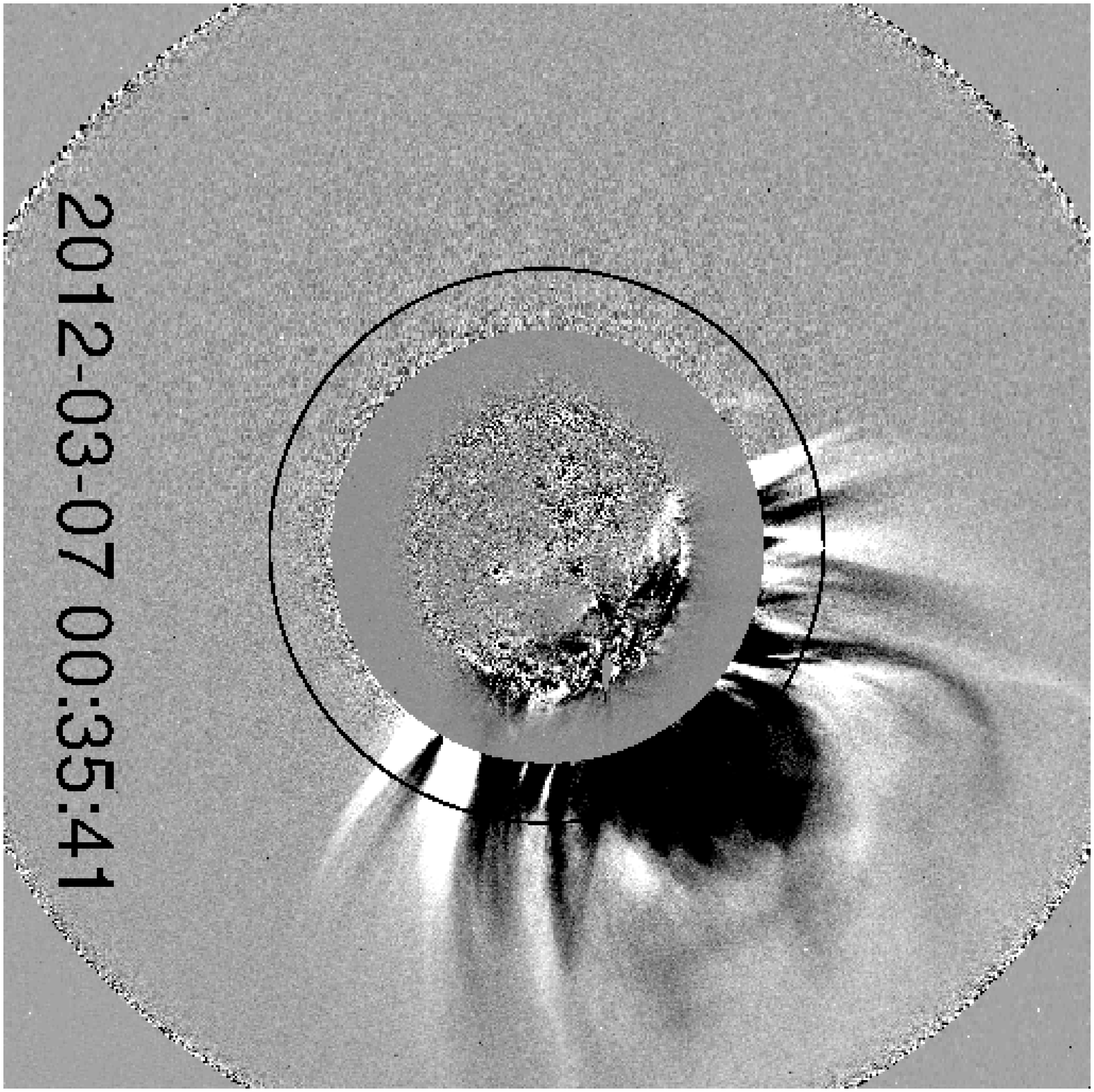}
\put(3,68){\color{black}{ \fontsize{4}{5}\selectfont CD 1}}
\put(17,70){\color{black}\vector(1,0){25}}
\put(8,58){\color{black}{ \fontsize{4}{5}\selectfont EUV 1}}
\put(22,60){\color{black}\vector(1,0){25}}
\put(7,88){\color{black}{ \fontsize{4}{5}\selectfont Traced Path 2}}
\put(20,86){\color{black}\vector(1,-1){15}}
\end{overpic}

\begin{overpic}[scale=0.01,angle=90,width=.6\textwidth,height=5.4cm,keepaspectratio]{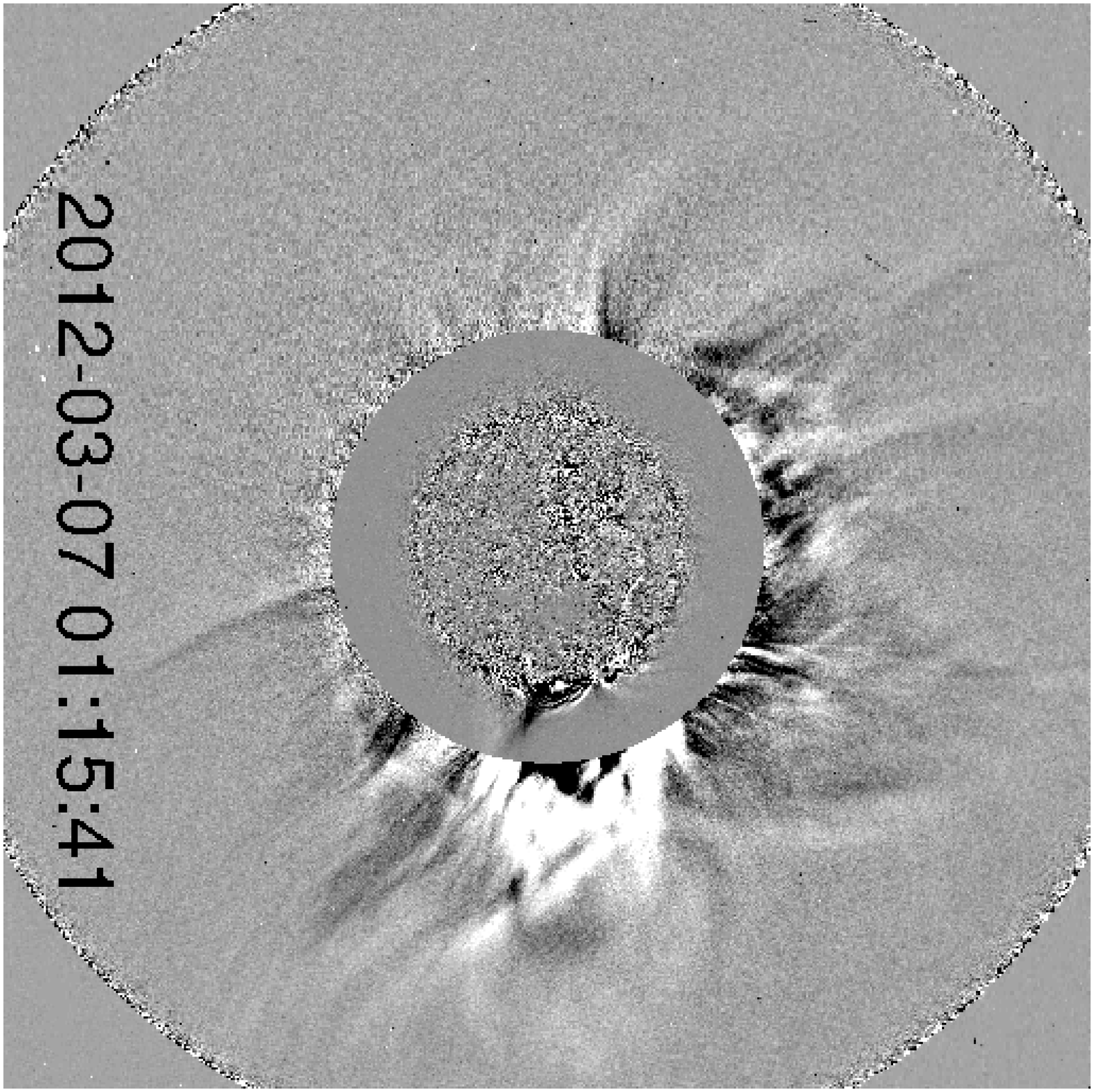}
\put(14,34){\color{black}{ \fontsize{4}{5}\selectfont CD 1}}
\put(20,40){\color{black}\vector(0,1){15}}
\put(25,82){\color{black}{ \fontsize{4}{5}\selectfont EUV 2}}
\put(36,80){\color{black}\vector(1,-1){25}}
\put(63,89){\color{black}{ \fontsize{4}{5}\selectfont CD 2}}
\put(69,88){\color{black}\vector(0,-1){25}}
\end{overpic}
\end{tabular}
\label{1}
\end{figure*}

Magnetohydrodynamic wave modes are ubiquitous in the solar coronal structures, which are potentially used in 
the diagnostics of the localized magnetic field and plasma properties in the solar atmosphere (e.g., Nakariakov \& Ofman 
2001; Srivastava et al., 2008; Van Doorsselaere et al. 2008; Srivastava \& Dwivedi, 2010; Aschwanden \& Schrijver, 
2011; White et al., 2012; Liu \& Ofman, 2014; Guo et al. 2015, and references cited there). MHD theory of magnetic flux tube and its 
wave modes have been established four decades ago, however, its continuous refinement provides a significant 
basis for making the inversion of observed wave and plasma properties and thus the potential use of the 
theory of MHD seismology in diagnosing solar corona (e.g., Uchida et al., 1979; Roberts et al., 1984; Nakariakov \& Verwichte, 2005;
Andries et al., 2005; McEwan et al., 2006; Verth et al., 2008; Macnamara \& Roberts, 2011; Luna-Cordozo et al., 2012; Li et al, 2015, and references cited therein). Recently, it has been found that realistic MHD models are required for more accurate coronal seismology related measurements where realistic magnetic field and initial atmospheric conditions have typical influence on the properties of evolved MHD waves. For instance, measured parameters (e.g., magnetic field) show deviation from the one estimated by classical MHD seismology (e.g., De Moortel \& Pascoe, 2009; Ofman et al., 2015). 

Extreme ultraviolet (EUV) waves are one of the most significant phenomena discovered by the EUV Imaging Telescope
(EIT, Delaboudini\'ere et  al., 1995) onboard the Solar and Heliospheric Observatory (SoHO). These waves are described 
in terms of various wave and non-wave models, e.g., the fast magnetoacoustic wave models (Ofman \& Thompson, 2002), stationary wave fronts on QSLs (Delann\'ee and Aulanier, 1999), flux-rope stretching model (Chen et al., 2002, 2005), successive reconnection model (van Driel-Gesztelyi et  al., 2008), slow-mode wave model (Wills-Davey \& Attrill, 2009), and the current shell model (Delann\'ee et  al., 2008). The flux-rope stretching model also describes the presence of fast mode waves ahead of it (Chen et al., 2002). It has recently been found that even the fast component of the EIT wave can be stationary at the QSLs (Chandra et al., 2016), which is also evident in the numerical simulation by Chen  et al. (2016). Obviously, EUV waves do interact with the localized magnetic structures to exchange the energy. The large-scale wave fronts can propagate across (or at any angle) w.r.t the coronal magnetic field likewise fast magnetoacoustic waves, and show the reflections from magnetic field (thus Alfv\'en velocity) discontinuities (Ofman \& Thompson, 2002). They also generate MHD waves and oscillations in the magnetic fluxtubes during resonant interaction (e.g., Ballai et al., 2005; Asai et al., 2012; Srivastava \& Goossens, 2013; Guo et al., 2015, and references cited there). 

\begin{figure}
\caption{Position-angle (P.A.) and time map for COR-1 disturbances measured at 1.6 solar radii. This height and measurement 
is shown by a black circle as indicated by an arrow in Fig.~1. On PA-time map, the triangles along two curved paths show the tracked points 
of two EUV disturbances in the corona. Kinematics of the first (bottom-left) and second (bottom-right) EUV disturbance as seen in COR-1 at 1.6 solar radii are displayed from the origin, i.e., P.A.=0.}
\includegraphics[scale=0.4,angle=90,width=8.8cm,height=5.5cm,keepaspectratio]{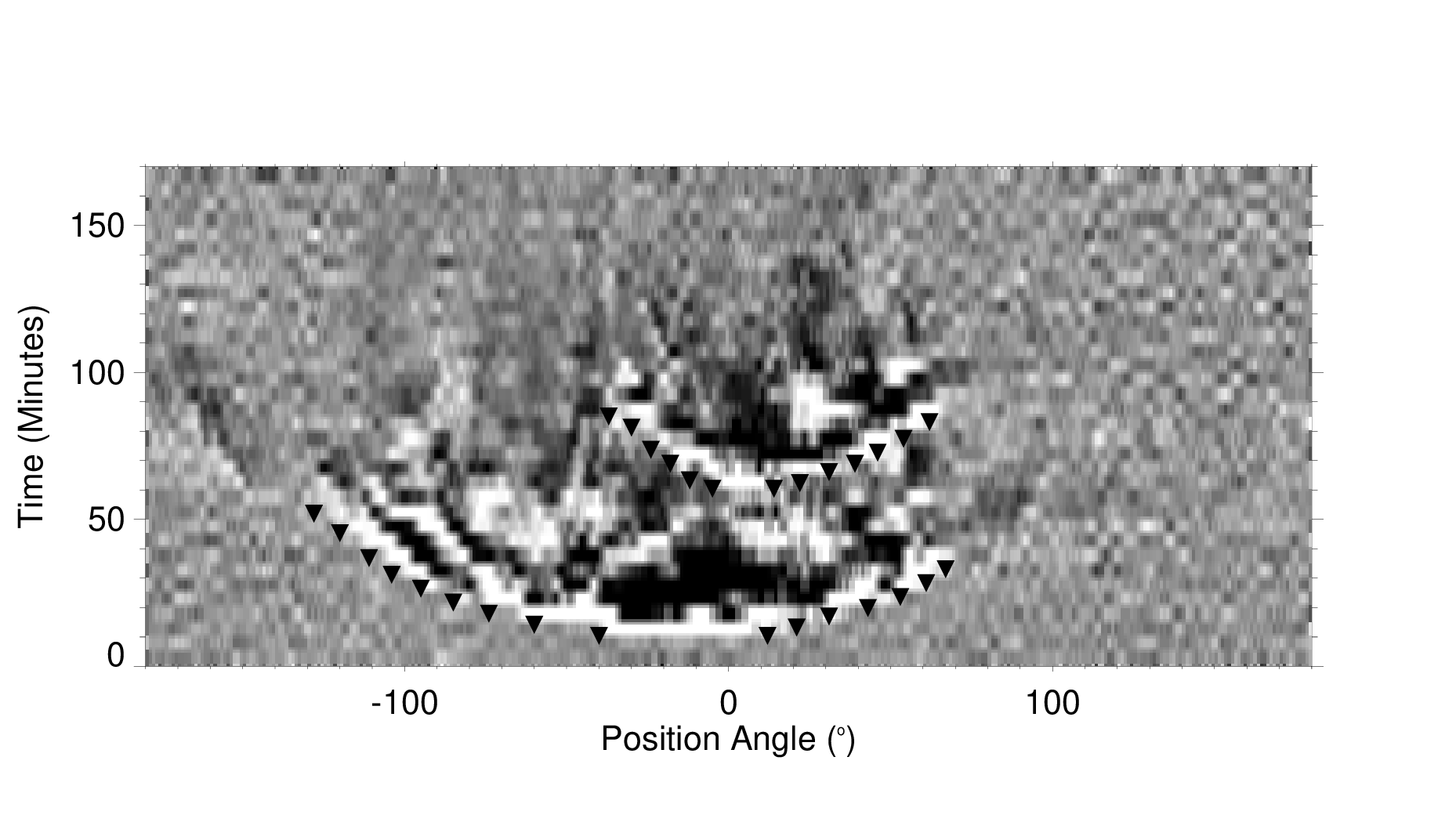}
\mbox{
\includegraphics[scale=0.2,angle=90,width=4cm,height=6cm,keepaspectratio]{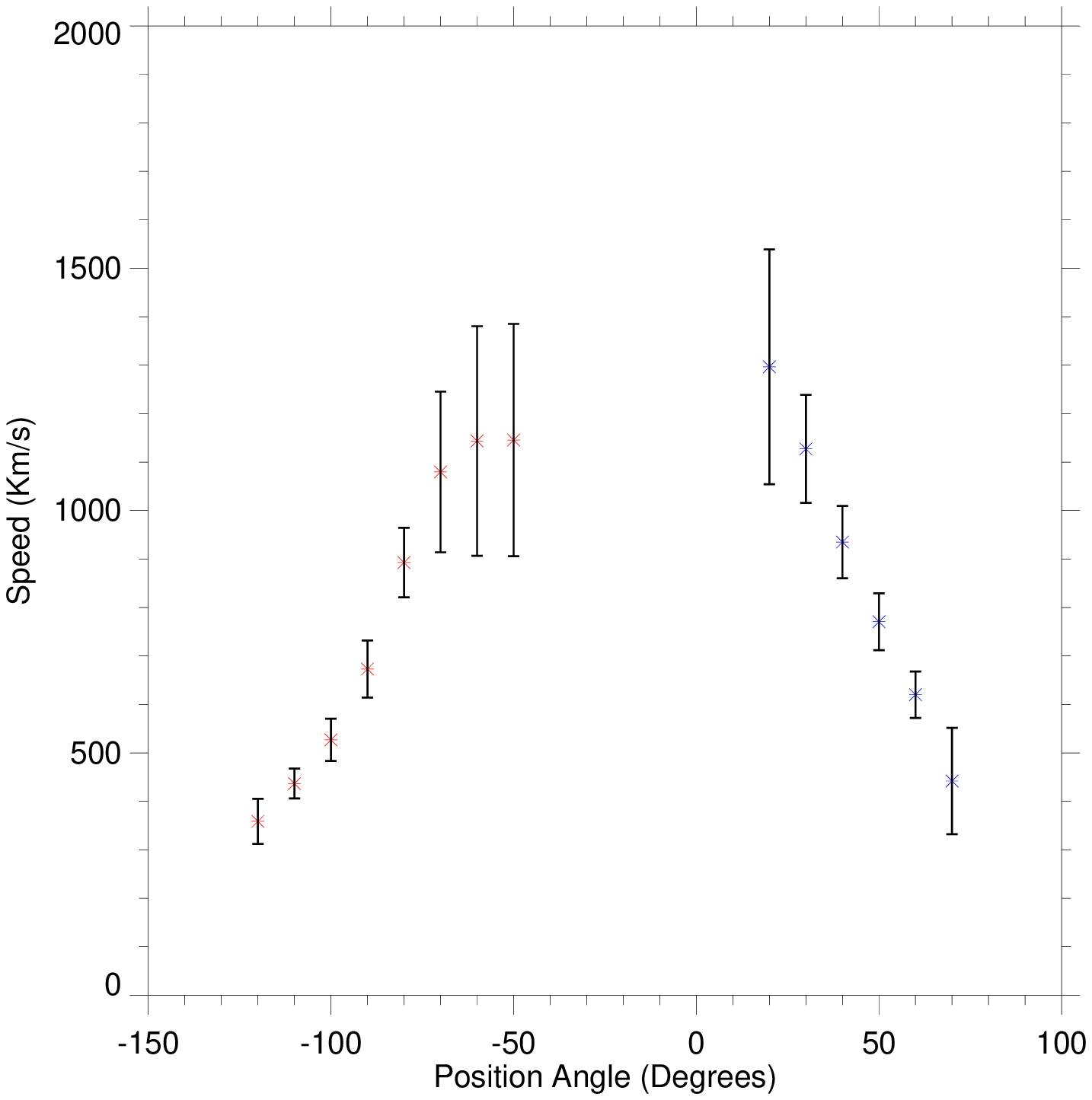}
\includegraphics[scale=0.2,angle=90,width=4cm,height=6cm,keepaspectratio]{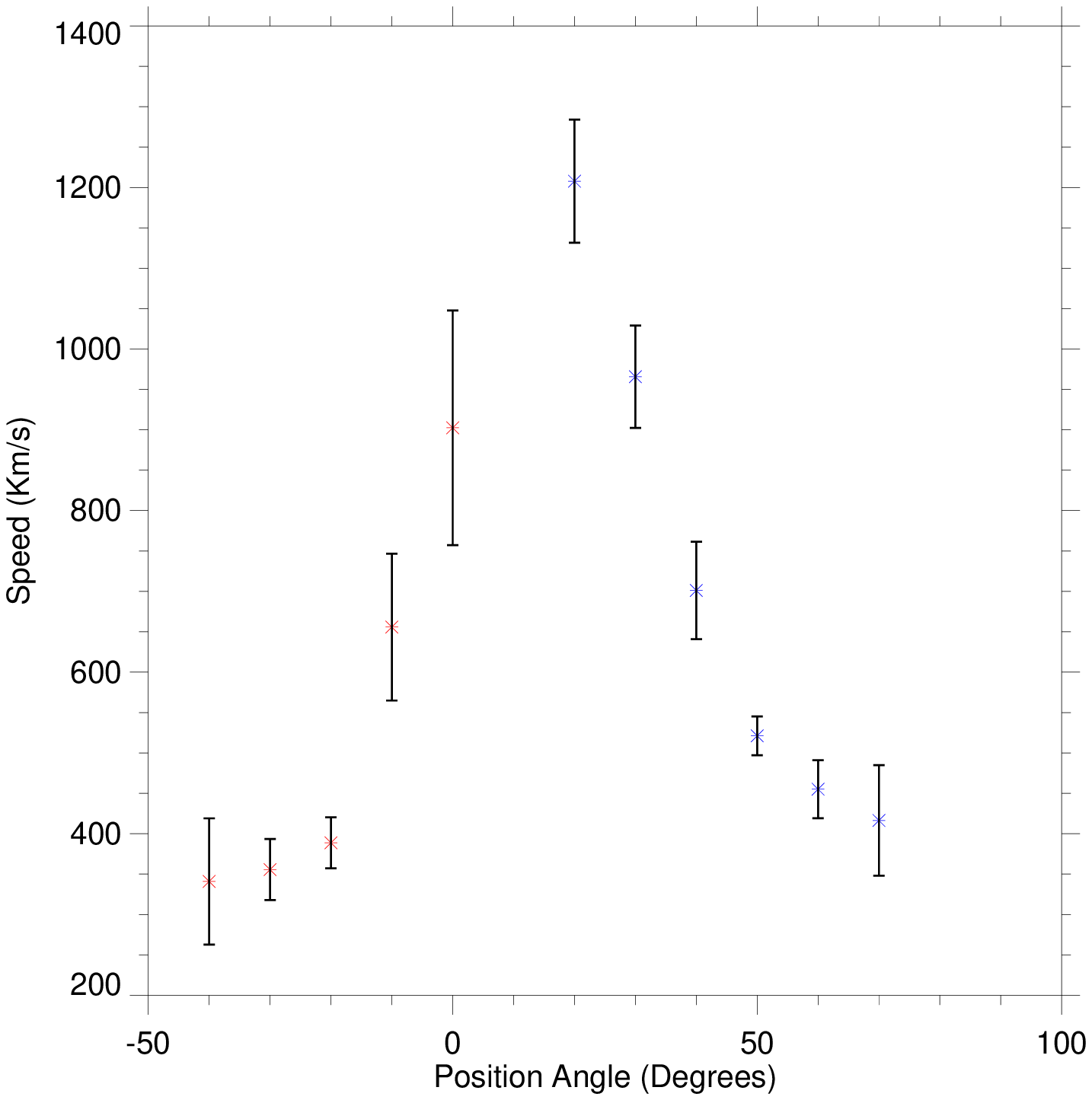}}
\label{2}
\end{figure}

The coronal streamers are the large-scale magnetic structures of the equatorial solar corona, which are deflected transversely during its interaction with EUV waves (Kwon et al., 2013). These transversal perturbations propagate in the form of kink waves to higher heights in the coronal streamer. These magnetic structures are basically associated with the slow solar wind in the near equatorial corona (see the review by Abbo et al., 2016). Therefore, understanding its physical conditions is of great importance in solar coronal physics. In the present paper, we describe the rare observations of the interaction of multiple EUV waves with a coronal streamer. The successive interactions result in two episodes of the transversal deflections of the streamer back-to-back generating kink waves propagating along it. We exploit the properties of these kink waves in the streamer tube to infer its magnetic field. Sect.~2 describes the Observational Data. In Sect.~3, we discuss the observational results and MHD seismology of the observed streamer. In the last section we outline the discussion and conclusions.

\begin{figure}
\caption{Top : Oscillating streamer as seen in STEREO-B/COR-1 on 7 March 2012 at 01:50 UT. This image is constructed by adding three polarization of 0, 120 and 240 degrees.
Bottom : Tracked streamer oscillations from 2.0 to 2.8 Solar radii. Horizontal axis is the time (in min) and vertical axis is normalized distance (Mm).}
\hspace{+0.7cm}
\includegraphics[scale=0.2,width=8.3cm,height=8.3cm,keepaspectratio]{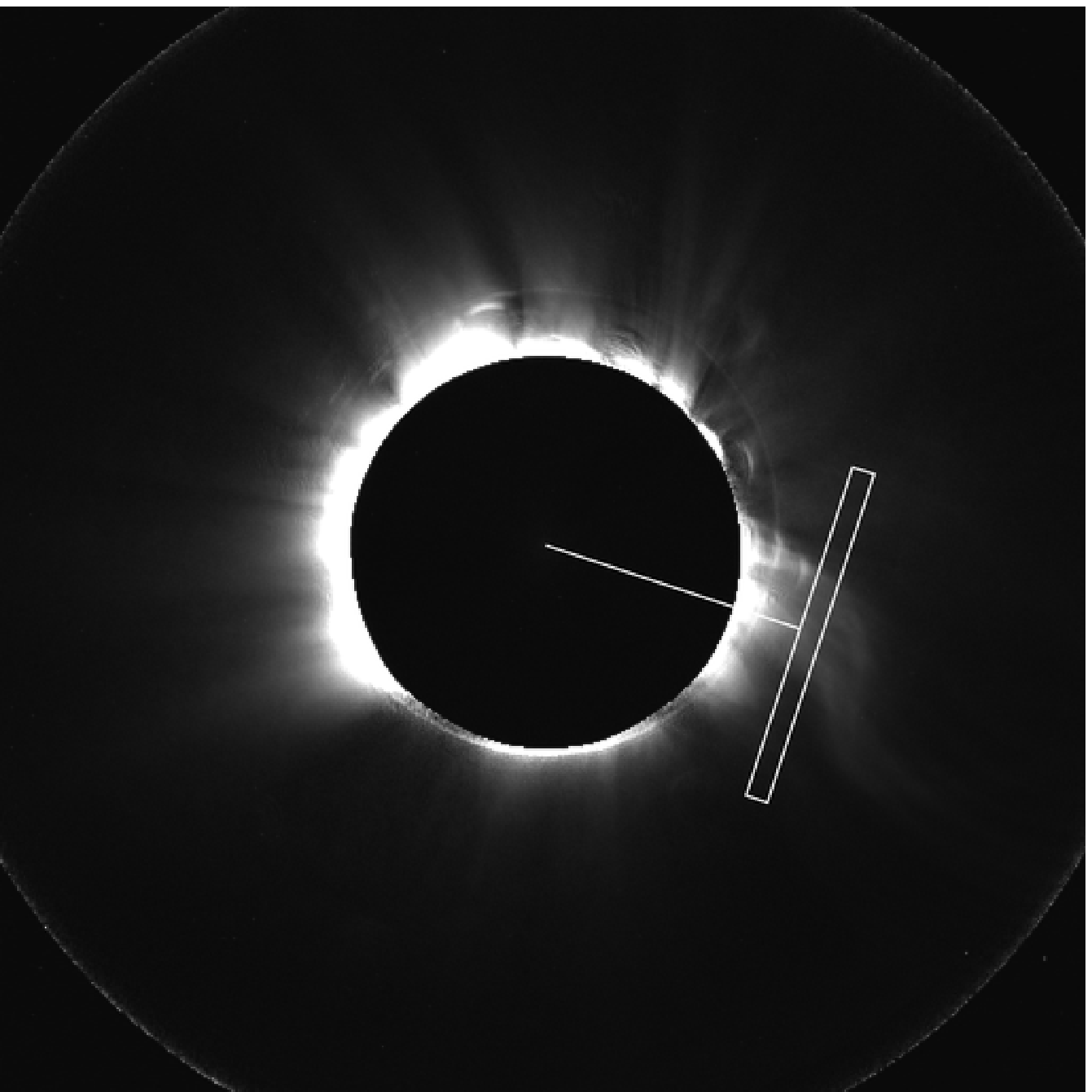}
\begin{tabular}{lll}
\begin{overpic}[scale=0.01,angle=90,width=.3\textwidth,height=3.2cm,keepaspectratio]{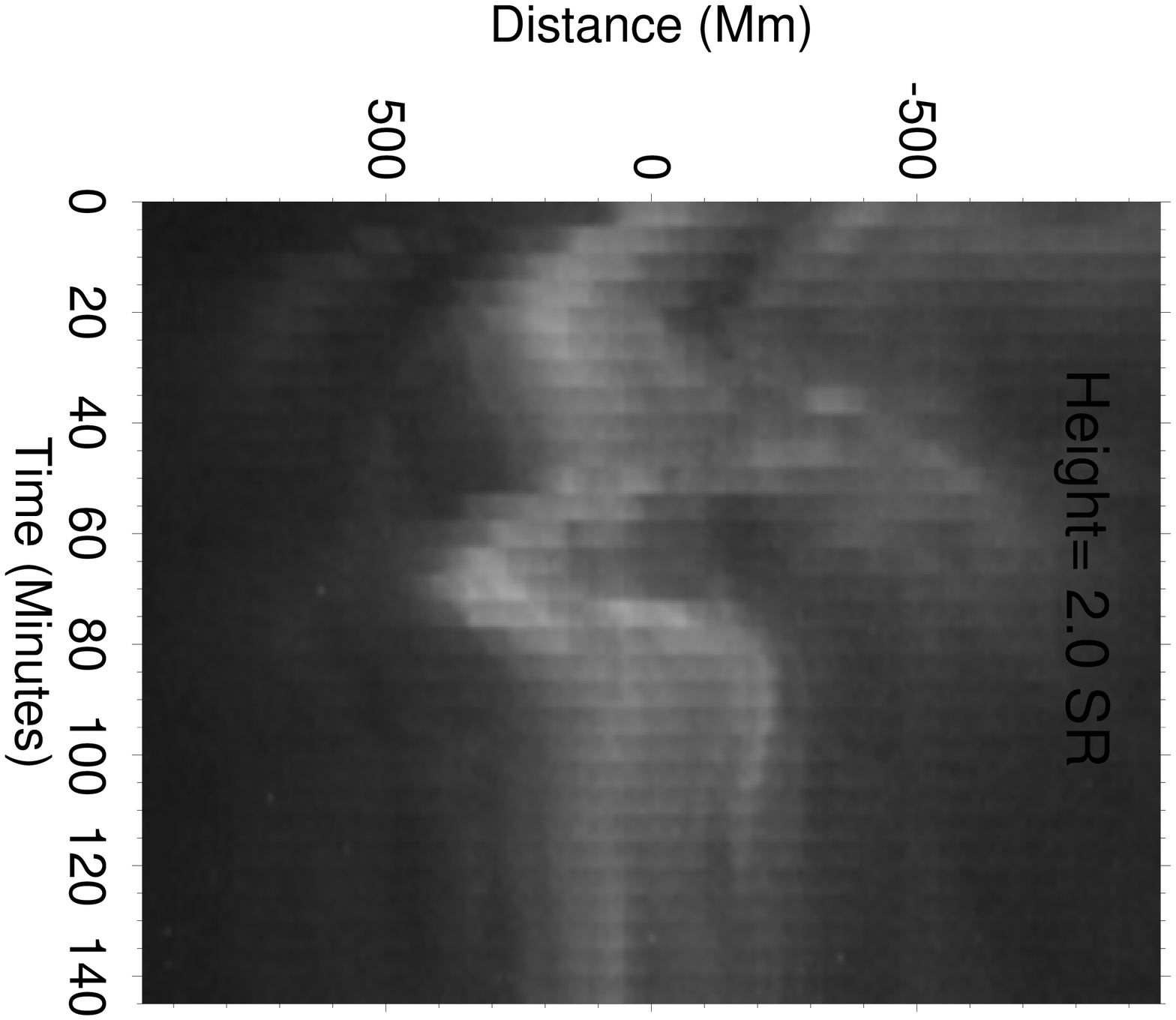}
\put(5.0,15){\color{red}{ \fontsize{5}{5}\selectfont Pulse 1}}
\put(12.5,20){\color{red}\vector(0,1){15}}
\put(33,10){\color{red}{ \fontsize{5}{5}\selectfont Pulse 2}}
\put(42,15){\color{red}\vector(0,1){15}}
\end{overpic}&
\includegraphics[scale=0.2,angle=90,width=3.2cm,height=3.2cm,keepaspectratio]{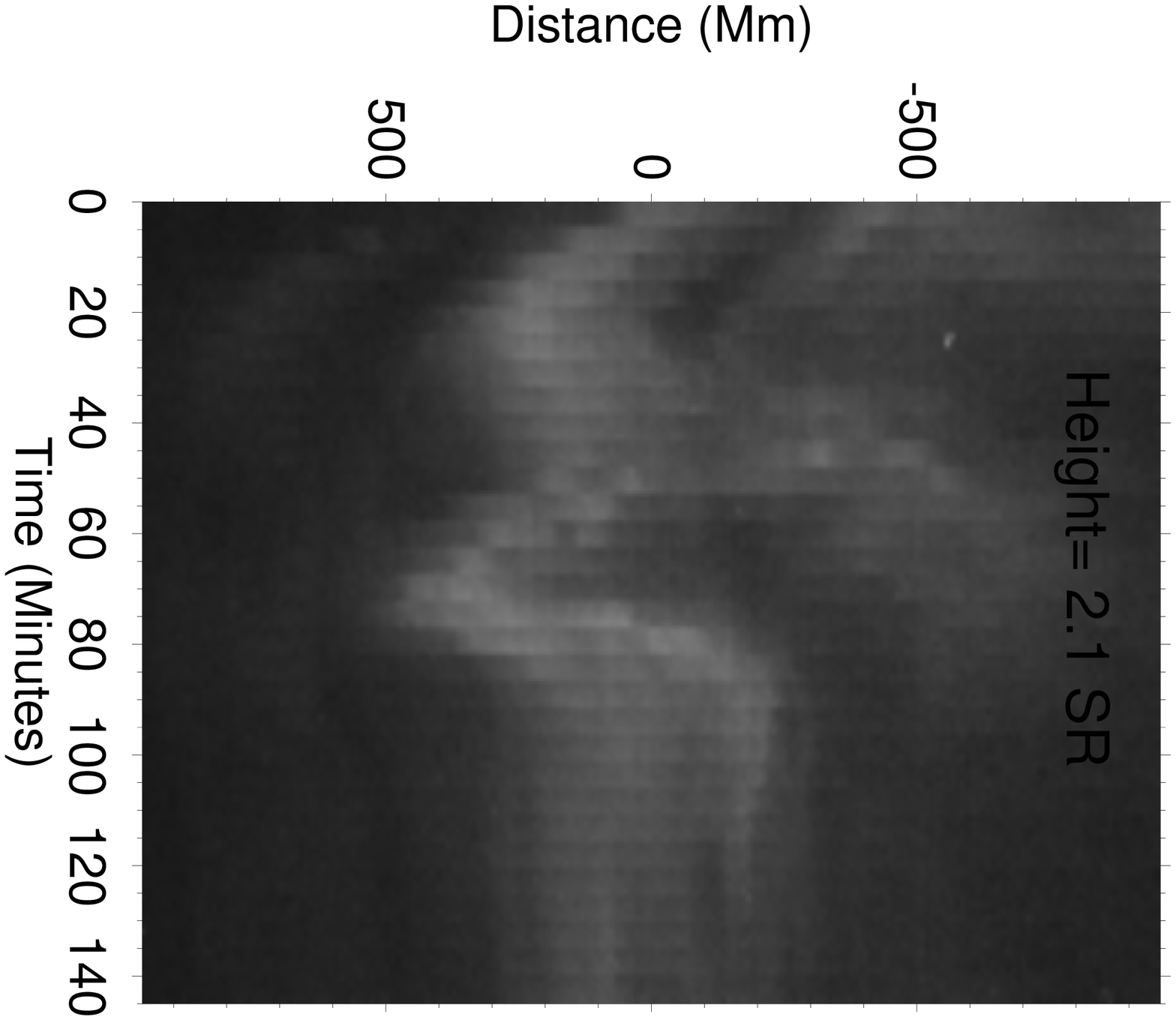}&\includegraphics[scale=0.2,angle=90,width=3.2cm,height=3.2cm,keepaspectratio]{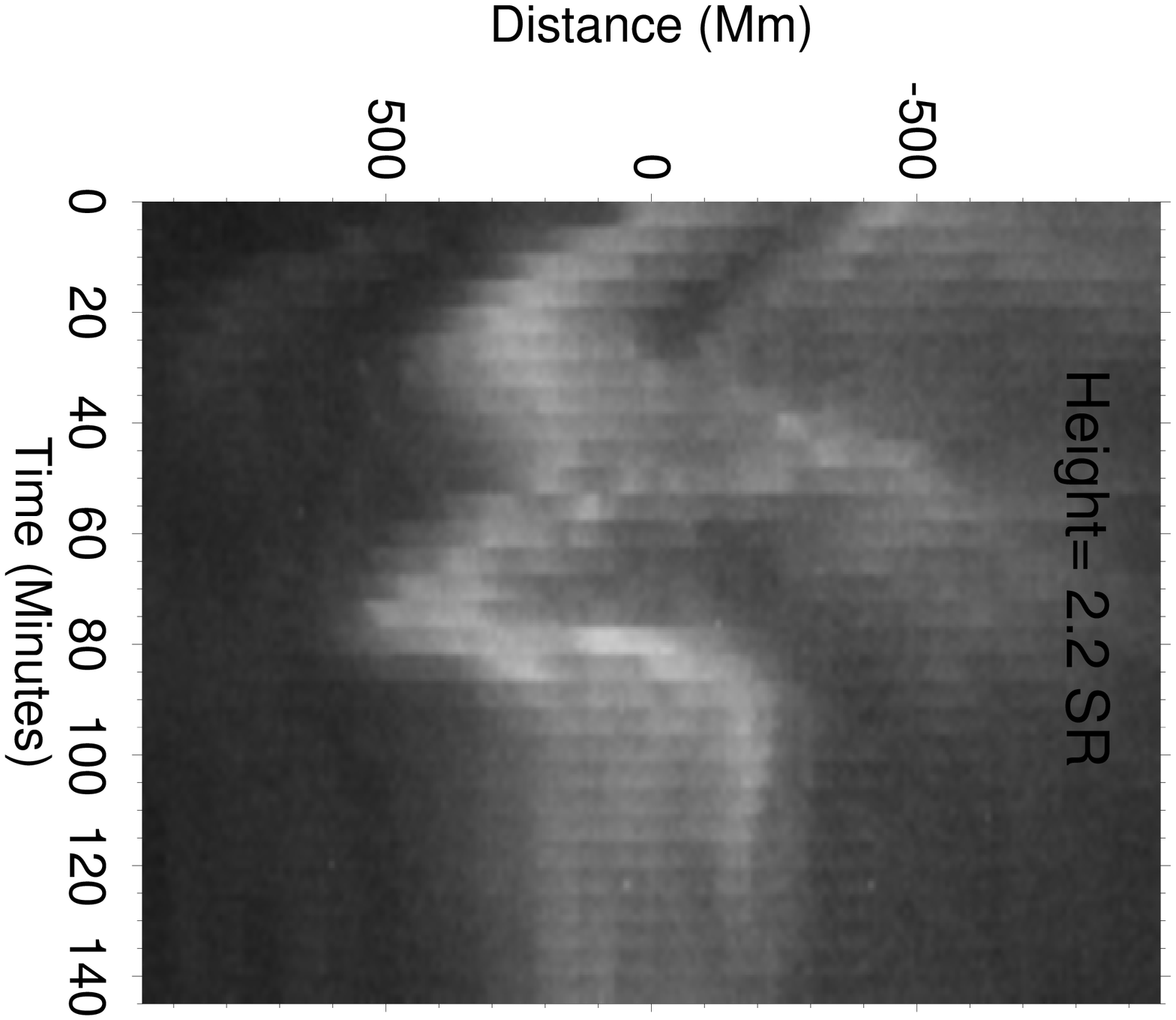} \\
\includegraphics[scale=0.2,angle=90,width=3.2cm,height=3.2cm,keepaspectratio]{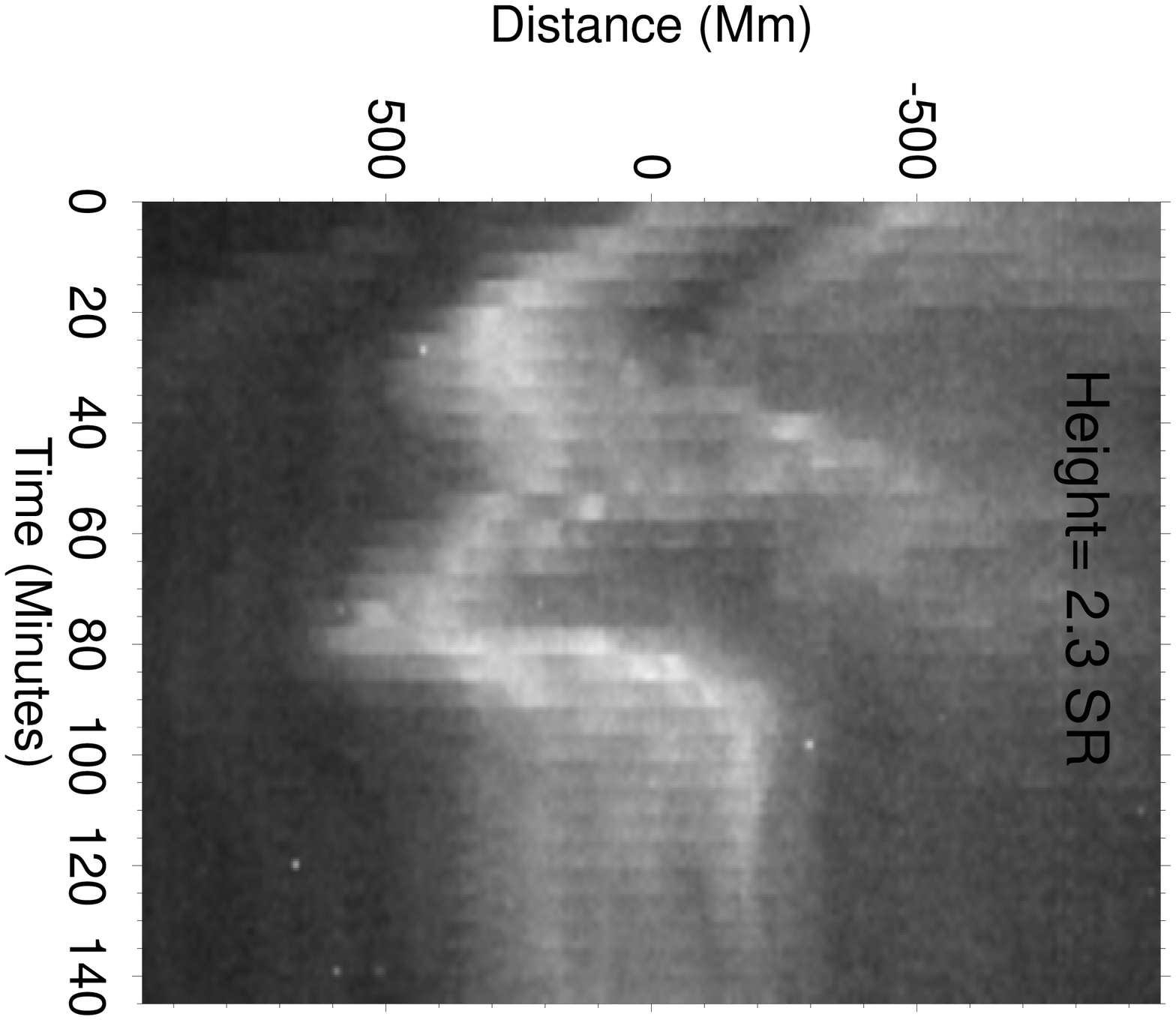}&\includegraphics[scale=0.2,angle=90,width=3.2cm,height=3.2cm,keepaspectratio]{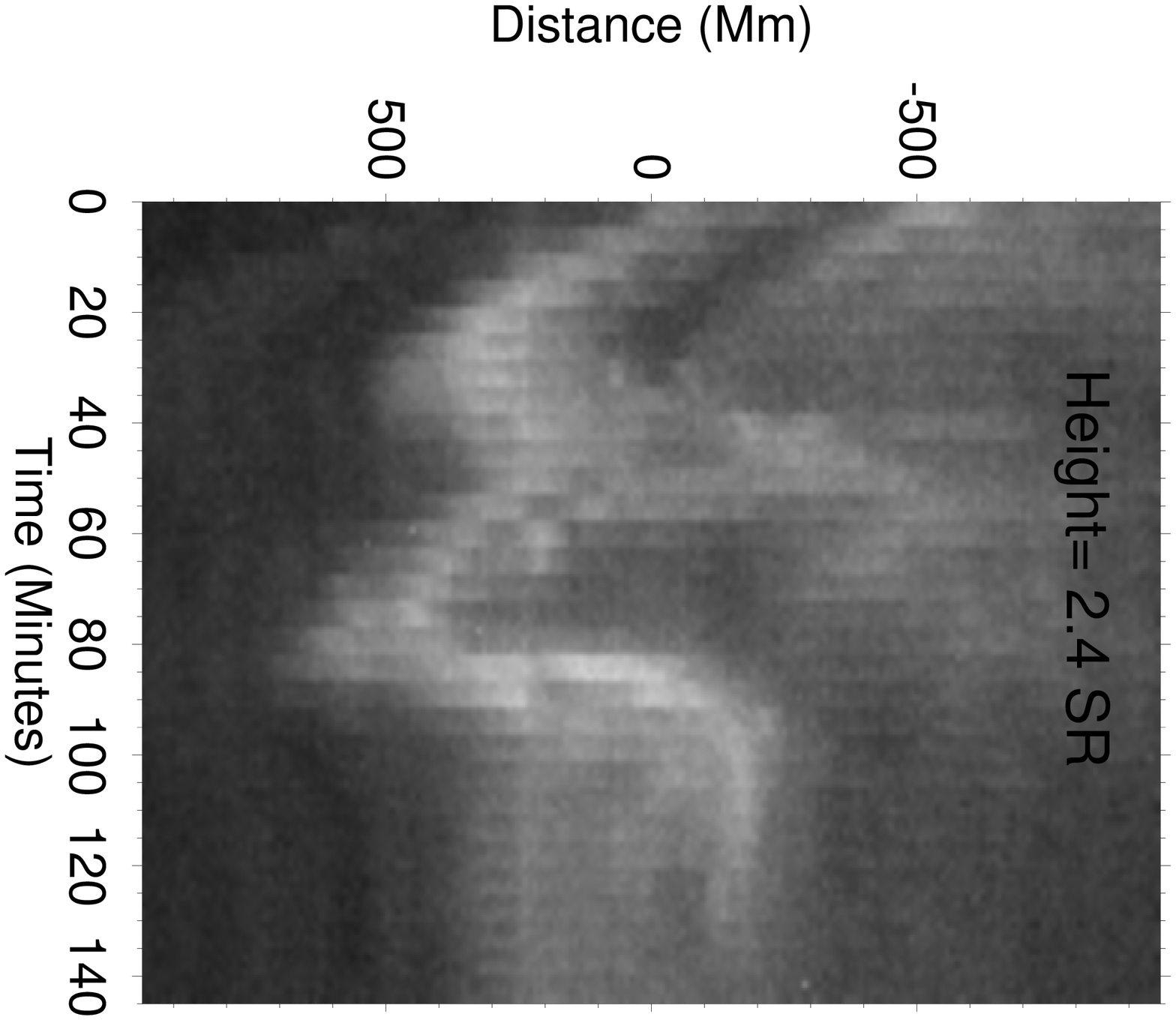}&\includegraphics[scale=0.2,angle=90,width=3.2cm,height=3.2cm,keepaspectratio]{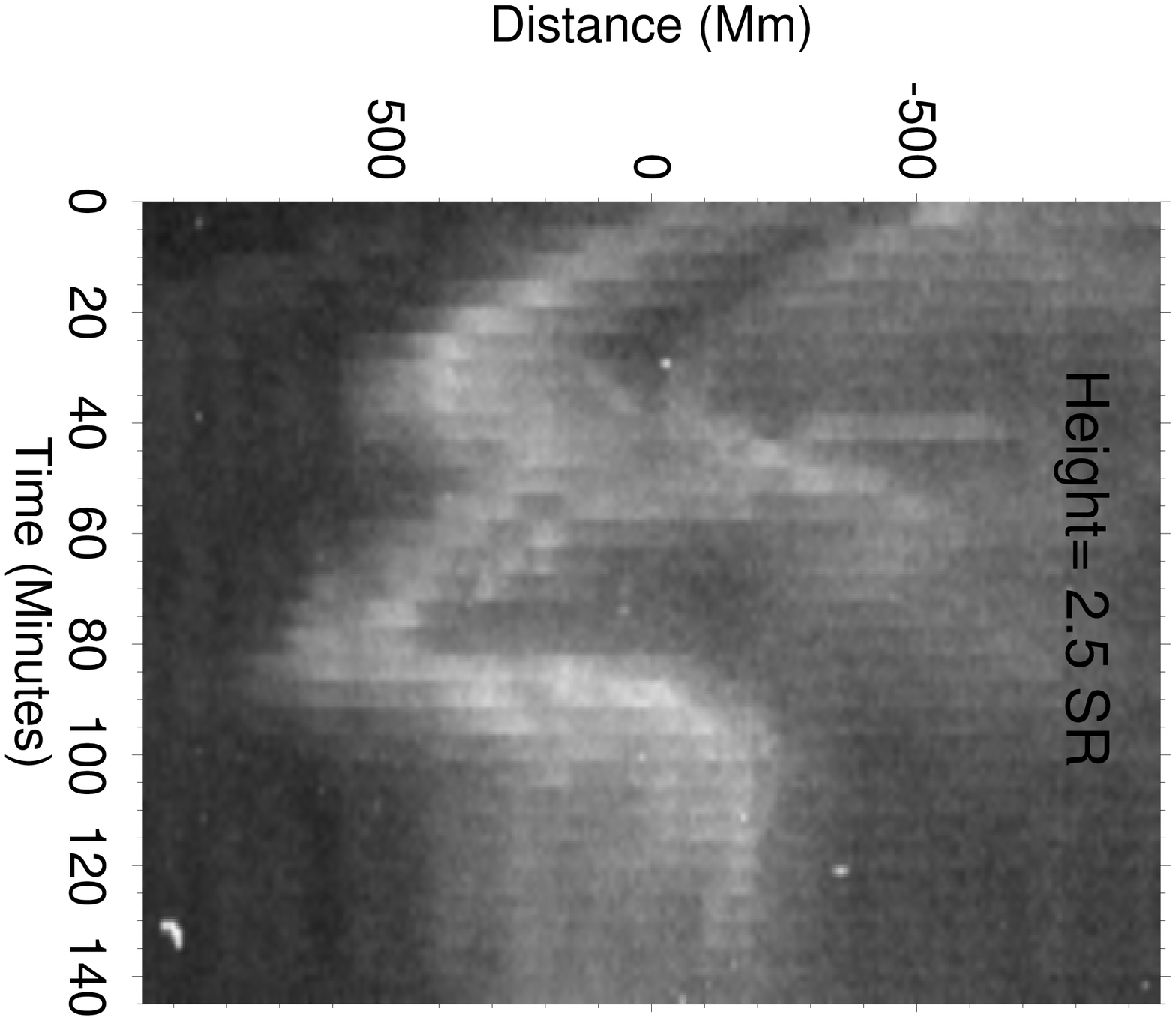}\\
\includegraphics[scale=0.2,angle=90,width=3.2cm,height=3.2cm,keepaspectratio]{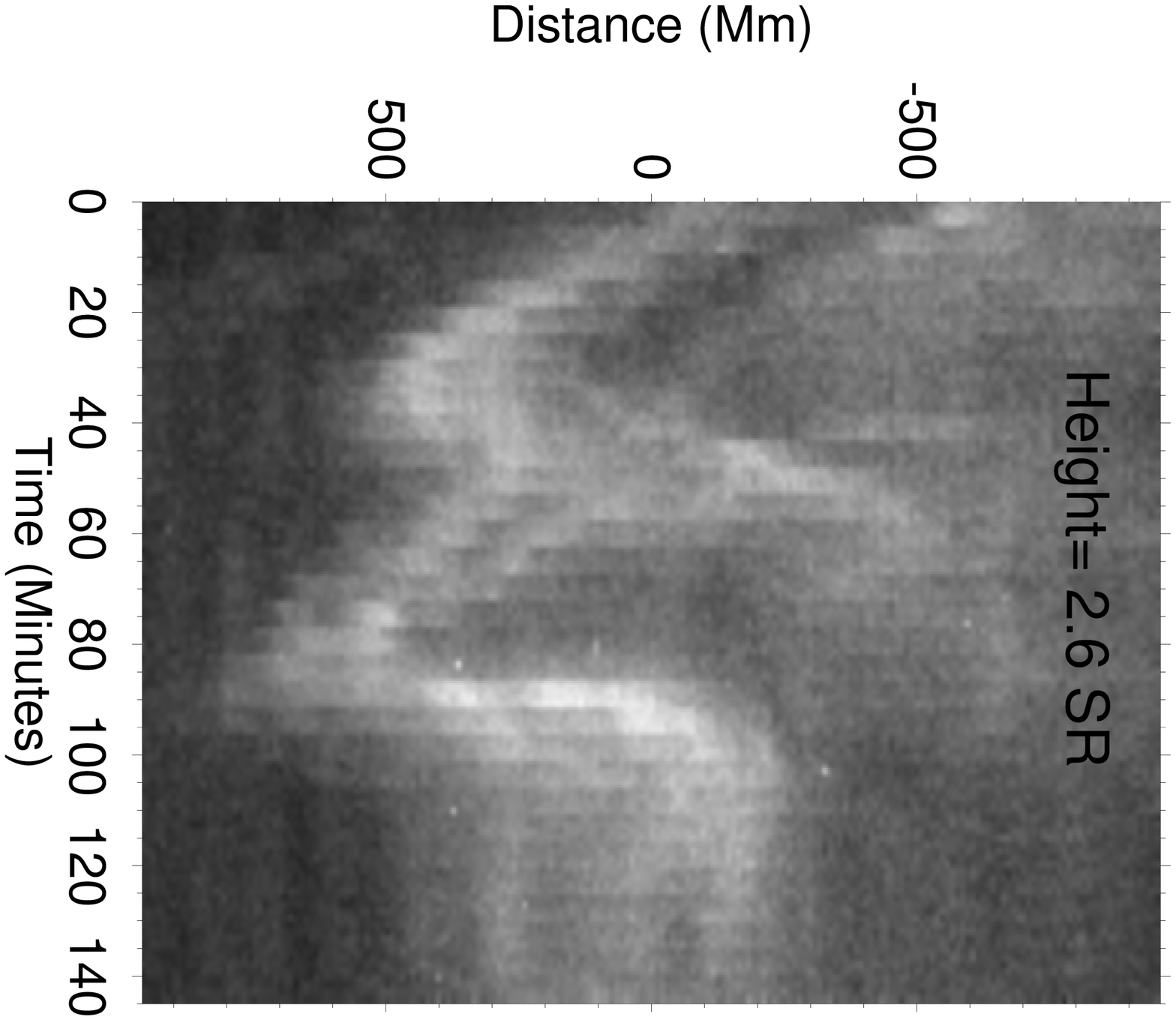}&\includegraphics[scale=0.2,angle=90,width=3.2cm,height=3.2cm,keepaspectratio]{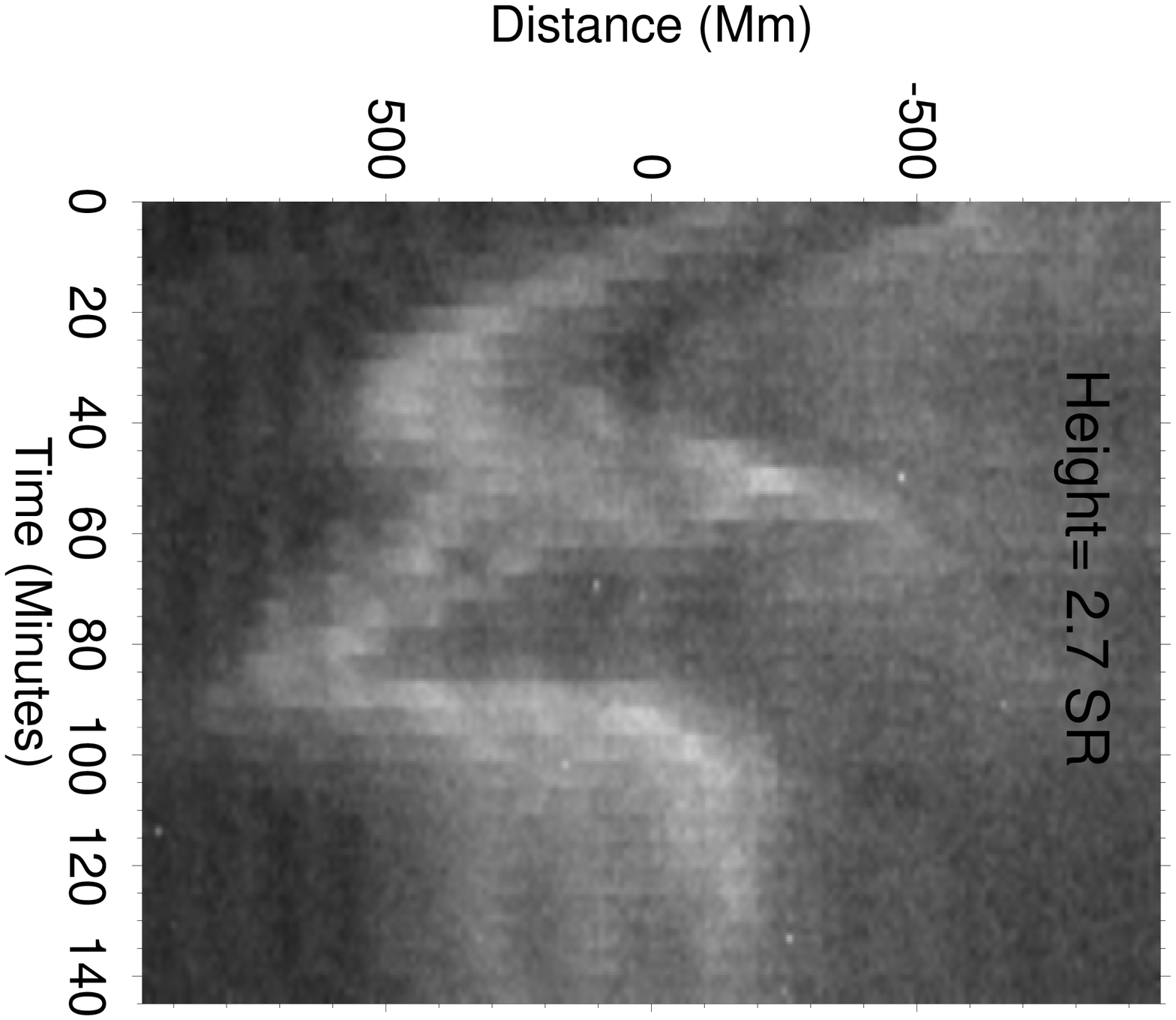}&\includegraphics[scale=0.2,angle=90,width=3.2cm,height=3.2cm,keepaspectratio]{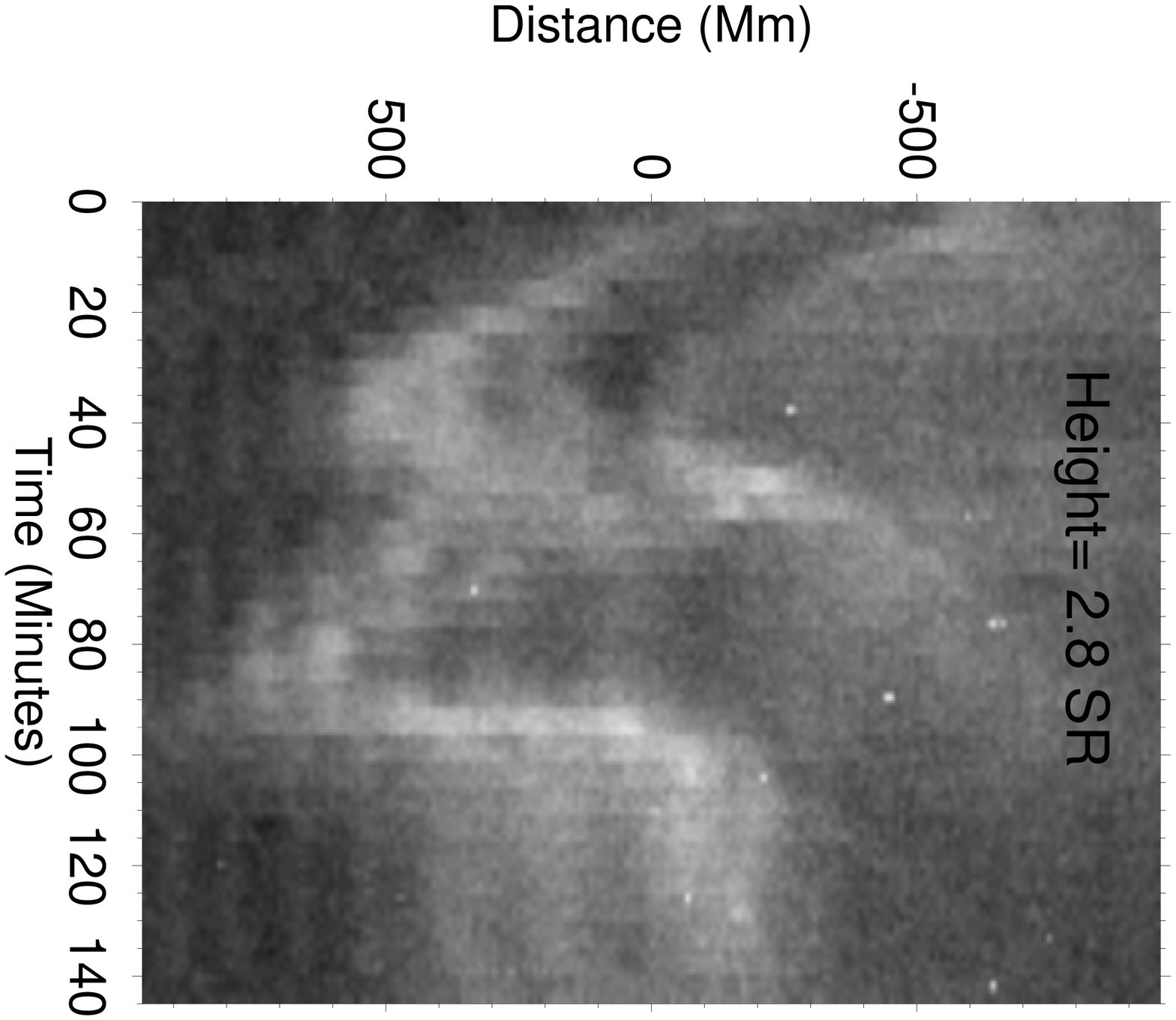}
\end{tabular}
\label{6}
\end{figure}
\section{Observational Data and Its Analyses}

In the present work, we observe two EUV waves using observational data from STEREO-B spacecraft. First and second EUV waves are triggered respectively at 00:05 UT and 01:05 UT on 7 March 2012. Observations of EUV wave propagation in the lower corona are seen in 195 \r{A} EUV data from Extreme Ultraviolet Imager (EUVI). To understand the upper coronal counterpart of these EUV waves as well as streamer oscillations, we use COR-1 coronagraphic observational data. COR-1 has a field of view of 1.5 to 4.0 solar radii. Both these instruments have time cadence of 5 minutes. EUVI and COR-1 instruments on board STEREO-B have a spatial resolution of 2.2 Mm and 23 Mm respectively for 2 pixels. STEREO data has been downloaded from "\path{http://secchi.nrl.navy.mil/cgi-bin/swdbi/secchi_flight/images/form}" and basic calibrations and background removal of data is done using Solarsoft IDL program "secchi\_prep.pro" and associated sub-routines. The details of STEREO spacecraft, and its instruments, i.e, EUVI \& COR-1, are given in detail by Howard et al. (2008) and Wuelser et al. (2004).

\section{Observational Results}

\subsection{Kinematics of EUV Wave Fronts in Outer Corona}
Fig. \ref{1} shows the running difference images for both EUVI and COR-1 data. EUV wave fronts are clearly visible in these snapshots. Top three image panels show the propagation of EUV wave fronts as seen by EUVI instrument in 195 \r{A} wavelength. Overlaying on the solar-disk is a curved path that is the path in most visible expansion of both the EUV wave-fronts. 
Along this path the first EUV disturbance shows the propagation speed of approximately 900 km s$^{-1}$ in the inner corona near its origin. The EUV disturbance at 1.6 solar radii possess enhanced speed near its origin which is approximately 1200 km s$^{-1}$ (cf., bottom-left panel of Fig.~2). This amplification is obvious with height in the stratified atmosphere. However, at a given height in the outer corona, both the EUV wave fronts decelerate with position angle (P.A.) on both the sides of their origin. We illustrate it in more detail below.

We basically aim to understand the kinematics of both the EUV wave fronts in the outer corona as seen in the STEREO-B/COR-1 observations. We choose a circular path at 1.6 solar radii (see Fig.~1 as indicated by black arrow) to capture the expansion of both the EUV wave-fronts. By stretching this path into a straight line and stacking horizontally with time, we get the Time-position angle plots.
This approach is used to get the kinematics properties of EUV wave disturbances as seen in COR-1. As shown in the bottom-middle image of Fig.~1, we take the pixels of a circular path at a particular height (here represented by $"$Traced Path 2$"$). We then straighten this circular path to a line and stack with all the lines in a time series horizontally. The result is a Time-Position angle (P.A.) plot as shown in top panel of Fig.~2. Two distinct EUV wave-fronts can be seen in this plot. We traced these fronts (black triangles). After converting PA-Time data to Distance-Time data, we found the speeds of these wave fronts by using three-point (quadratic) Lagrangian interpolation. The results have been reported in bottom-left and bottom-right panels of Fig.~2. Both waves were decelerating in nature with different initial speeds as they spread in the corona from their origin on both the sides. The wavefronts of both the EUV waves, which propagate towards the coronal streamer situated adjacent in right hand side from their origin site, decelerate. This qualitatively suggests their interaction with the coronal streamer and other localized magnetic fluxtubes in their path and thus the energy exchange (see Murawski et al., 2001; Ballai et al., 2005). Such energy exchange may cause the resonant MHD oscillations in the confined magnetic fluxtubes (e.g., streamer here), and generates the localized tubular MHD modes, i.e., kink wave in the present case. 

\begin{figure*}
\caption{Top-left: Phase-speed of kink waves generated in the streamer by first EUV wave at its different heights. Top-right : Phase-speed of kink waves generated in the streamer by second EUV wave at its different heights.}
\mbox{
\includegraphics[scale=0.2,angle=90,width=7cm,height=7cm,keepaspectratio]{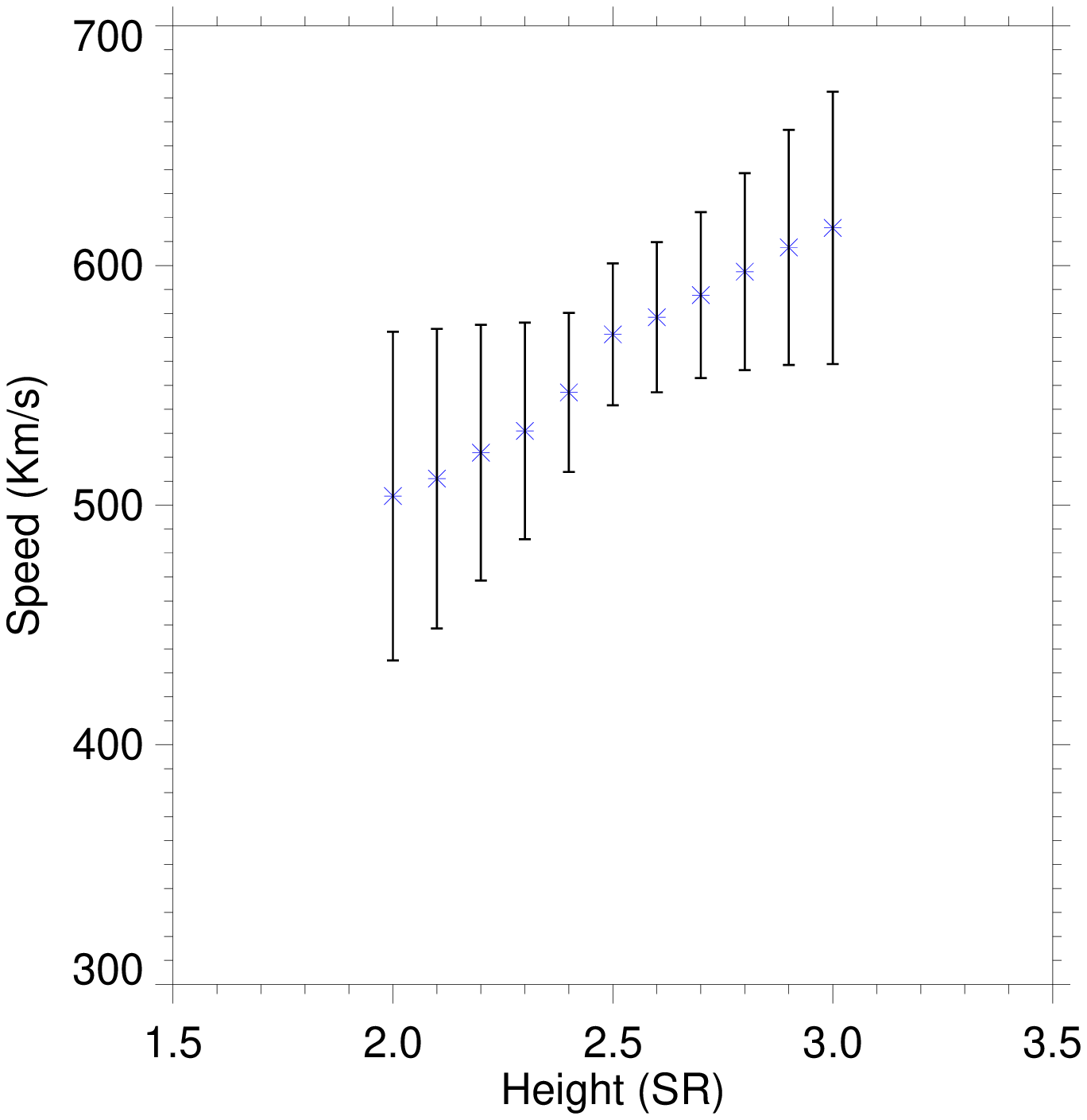}
\includegraphics[scale=0.2,angle=90,width=7cm,height=7cm,keepaspectratio]{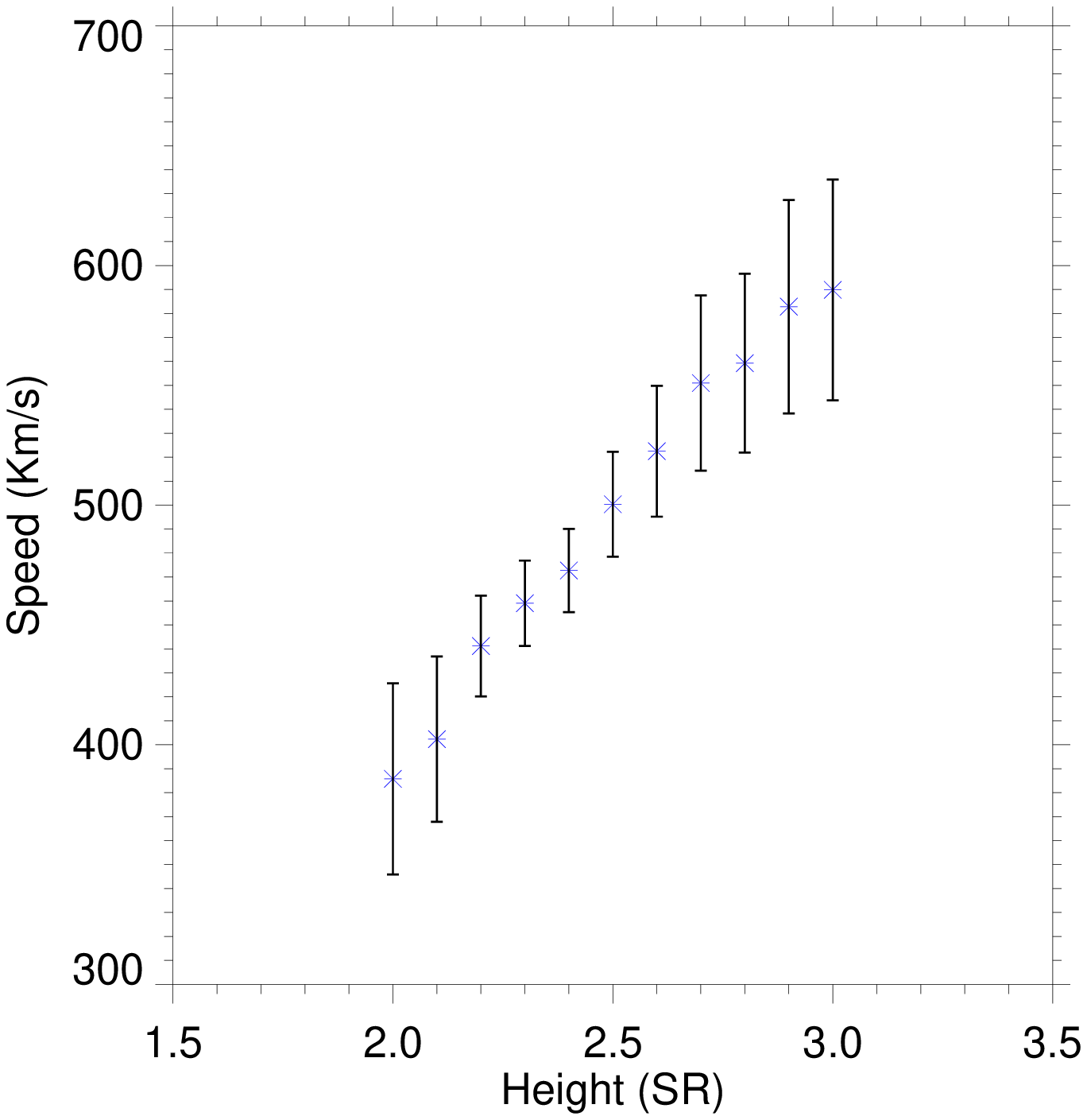}} 
\end{figure*}

\begin{figure*}
\caption{Left : Paths chosen along the streamer for SSPA inversion. Right : Density inside and outside the streamer as obtained by inversion of polarized brightness as observed by COR-1.} 
\mbox{
\includegraphics[scale=0.2,angle=0,width=7cm,height=7cm,keepaspectratio]{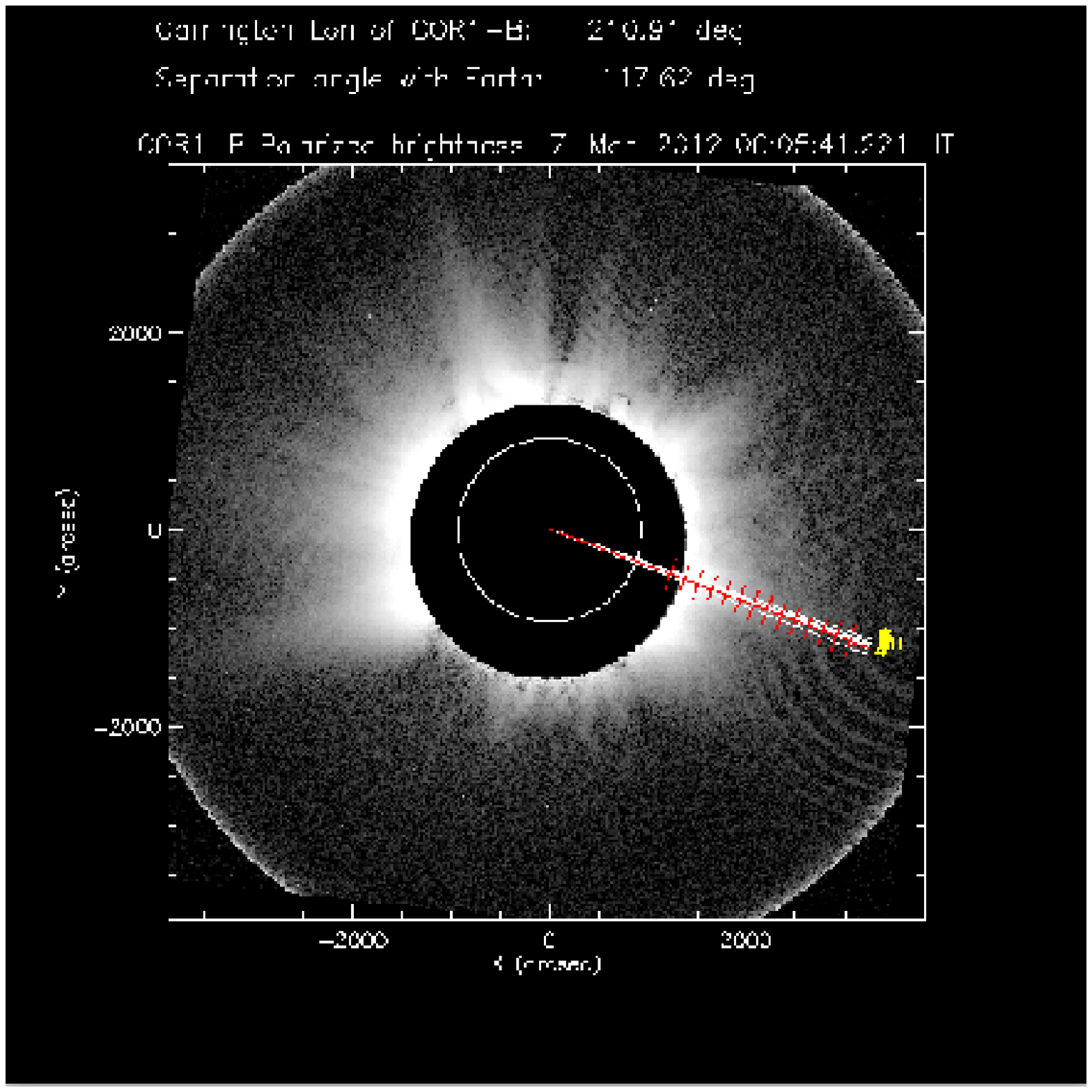}
\includegraphics[scale=0.2,angle=90,width=7cm,height=7cm,keepaspectratio]{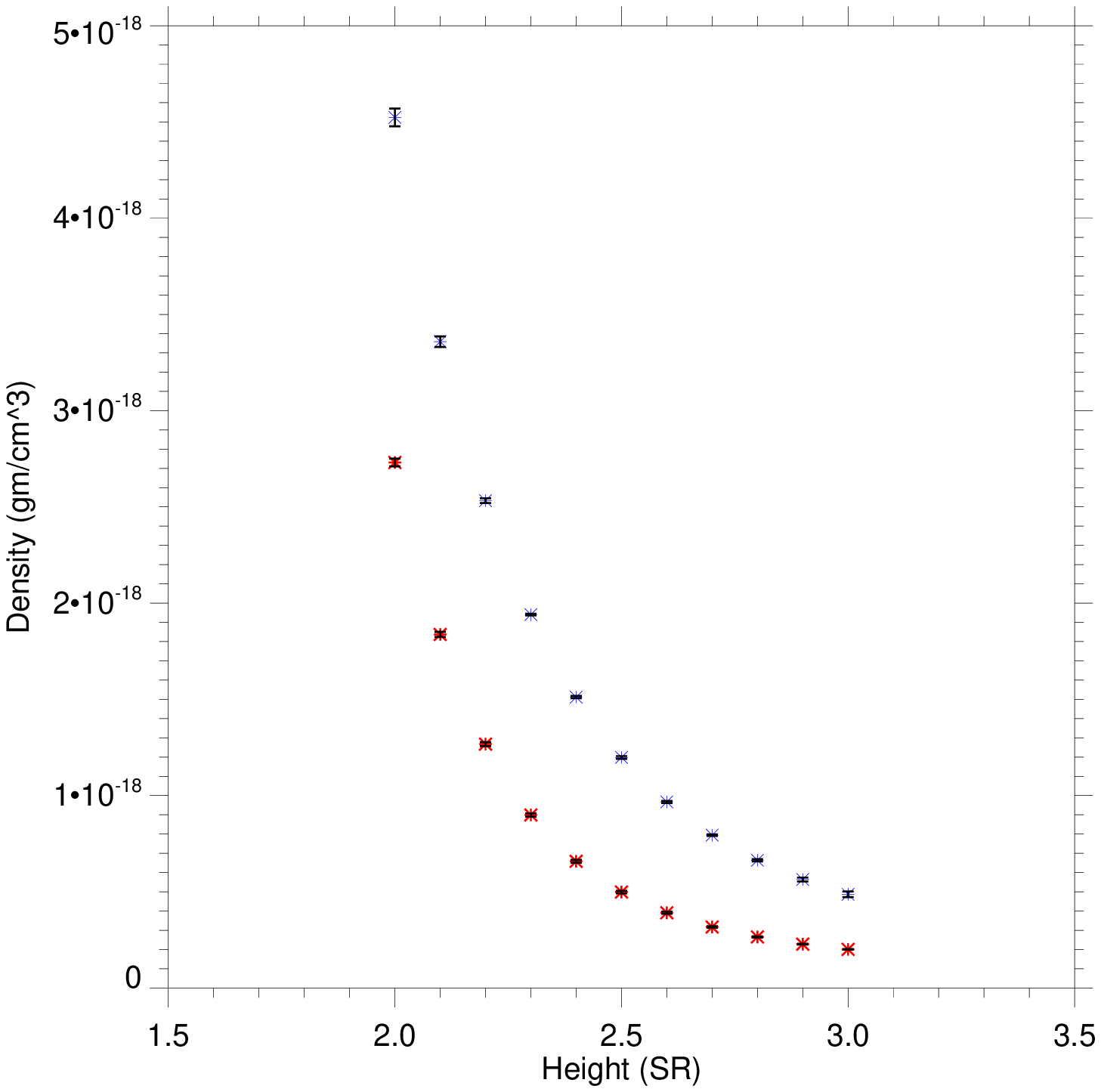}}
\end{figure*}

\subsection{Oscillations of the Coronal Streamer}

Fig.~3 is a white light image from COR-1 coronagraph formed by adding three polarization of 0, 120 and 240 degrees. This image is taken at 01:50 UT on 7 March 2012. We can see that the kink wave is triggered in the streamer by its interaction with EUV waves. To get the phase speed of this wave, we see the portion of streamer at different heights for a time period. We take time slices as shown in Fig.~3 (top-panel) from the COR-1 temporal image data and stack them in a distance-time map. Each row of the bottom panel shows such distance-time maps for various heights from 2.0 to 2.8 solar radii.

At a given particular height, when the first EUV wave strikes, it deflects the coronal streamer transversely. Streamer further tries to restore its transverse displacement towards its equilibrium position. We consider this motion due to the effect of any energetic pulse of certain duration (by EUV wave) that bends the axis of the streamer tube as a whole. We have examined it by fitting a sinusoidal function on it (not shown here), and the nature of this transverse oscillation is found to be similar as previously observed by Kwon et al. (2013). This propagates as a kink wave at various heights of the streamer. Similarly, when second EUV wave acts upon the coronal streamer, it generates kink perturbations propagating-up. In case of first transverse perturbation (due to first EUV wave) when it was restoring, the second perturbation (due to second EUV wave) does occur. Therefore, it appears like a wave train at different heights of the streamer in distance-time maps. 

For each transversal kink perturbation (first and second), we track it multiple times manually at a particular height in the streamer. We repeat this at different heights of streamer also. We find the standard deviation and thus the inherent error in the peak position of the displacement at various heights. In both the cases, we fit second order curves to this data. By differentiating the fitted function, we find the phase speed at these heights. Speed evolution of these two kink oscillations with height is shown in top-left and top-right panels of Fig.~4. Both these waves show increase in propagation speed with the height within the streamer. Similar methodology is also adopted in even small-scale fluxtubes (e.g., spicules) to measure the phase speed of kink waves with height (Zaqarashvili et al., 2007).

\subsection{Estimation of Average Electron Density in and outside the Streamer}

We  use polarized brightness images of COR-1 to obtain the electron densities in and outside of the streamer as a function of height. We use Spherically Symmetric Polynomial Approximation (SSPA) method to estimate the densities (Wang \& Davila, 2014). This method measures the polarized brightness (pB) from COR-1 observational data at defined points along a chosen path in the corona. Hayes et al. (2001) assumed the radial electron density distribution in the corona in polynomial form and utilized it to the inversion of total brightness observations.
Wang \& Davila (2014) have used this technique for the pB inversion. The density distribution is assumed to be in a polynomial form, but the expression of pB is not in the polynomial form, but pB is a linear function of the coefficients of polynomial.
They determine the coefficients of the polynomial by a
multivariate least-squares fit to the curve of pB. The radial distribution of electron
density is then obtained by substituting the resulting coefficients directly into
the polynomial form. They have performed some experiment using coronal density models for the region between 1.5 and 6.0 solar radii to select appropriate degrees of polynominal as outlined in Saito et al. (1977), and found that choosing the first five terms can determine radial electron density with the relative errors lying within 5\% and reproduce
the observed pB measurements with the relative errors within 1\%. Therefore, they use the 5-degree
polynomial fits for all the SSPA inversions in their study. This inversion method provides resonably small errors in the electron density estimation, and reproduce well the observed polarized brightness. However, it should be noted that due to the limitation that a spherically symmetric geometry is assumed for the analyzed coronal structure, the method may underestimate coronal density by 20\% to 40\% (e.g. for streamers).

To convert electron densities into mass density, we assume streamer plasma made-up of 90\% H atoms and 10\% He atoms by mass. The density profile as found by this method is shown in right panel of Fig.~5. The density inside the streamer is found to be approximately two times larger than outside ambient corona. Each SSPA inversion for the chosen path within the streamer measures the density at various heights along with it. The relative error estimation is mentioned above using SSPA technique. We have further chosen ten paths by slightly varying the position angle within the streamer (left panel of Fig.~5), and measured the ten values of the density at each height by SSPA inversion technique. The standard deviation is calculated on average electron density value at a given height in the streamer.

Between 2-3 solar radii in the coronal streamer, the mass density varies from $\approx$(4.5$\pm$0.05)$\times\,10^{-18}$ to $\approx$(4.9$\pm$0.1)$\times\,10^{-19}$ g cm$^{-3}$. Between 2-3 solar radii, the mass density varies outside the streamer from $\approx$(2.75$\pm$0.02)$\times\,10^{-18}$ to $\approx$(2.0$\pm$0.015)$\times\,10^{-19}$ g cm$^{-3}$. In the SSPA method, the polarized brightness inversion has minimal relative error less than 1\%. Therefore, the devised density in the corona though 3-D model provides its fairly accurate values with less uncertainities.  

\subsection{Inference of the Magnetic Field inside the Streamer}

We use the kink speed ($c_{k}$), density inside ($\rho_{o}$) and outside ($\rho_{e}$) the streamer to find the magnetic field strength within it at different heights. We assume streamer as a cylindrical coronal waveguide, and theory of kink waves in it (Roberts, 1984; Nakariakov \& Ofman, 2001). The equations that relate kink speed, densities, and Alfv\'en speed which in turn depends upon the magnetic field, are given as follows :

\begin{linenomath*}
\begin{equation}
\label{eq:model}c_k=c_A \sqrt{2\rho_o/(\rho_o+\rho_e)}
\end{equation}
\end{linenomath*}
\begin{linenomath*}
\begin{equation}
\label{eq:model}c_A=B_o/\sqrt{4\pi\rho_o}
\end{equation}
\end{linenomath*}

Here, $\rho_o$ and $\rho_e$ are the plasma densities inside and outside the streamer respectively, $c_k$ is the kink
speed,  $c_A$ is the Alfv\'en speed and $B_o$ is the magnetic field strength. Using these equations, we estimate the magnetic field strength in the streamer.

Diagnosed magnetic fields along the streamer as estimated using the principle of MHD seismology are shown in Fig.~6. First (blue) and second (red) profiles of streamer's magnetic field are diagnosed respectively by using first (during first impact of EUV wave) and second (during second impact of EUV wave) kink wave motions in the streamer. It should be noted that both the EUV waves are well distinguished in time with an approximate difference of one hour. Both the magnetic field profiles show the exponential decay with a radial dependence, i.e.,  B$_{s}$=$A$$\,r^{l}$. Here $A$=0.54, 0.29 and $l$=-2.40, -1.84 respectively for first and second magnetic field profiles.

\begin{figure}
\caption{Diagnosed magnetic field profiles along the streamer using principle of MHD seismology.}
\includegraphics[scale=0.2,angle=90,width=7cm,height=7cm,keepaspectratio]{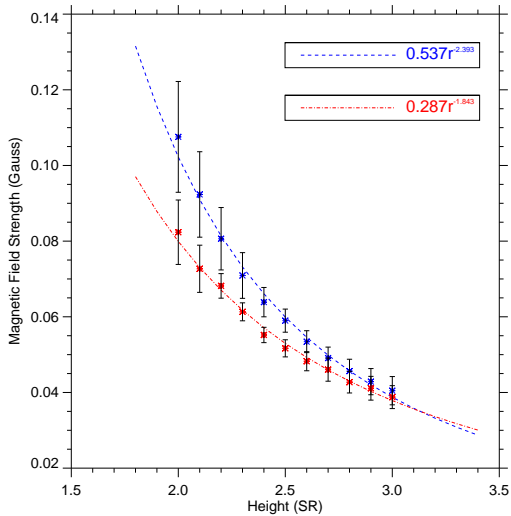}
\end{figure}

\section{Discussion and Conclusions}

Estimation of the streamer's magnetic field profile observed with STEREO/COR-1 on 7 March 2012 is performed using two kink wave episodes when two consecutive EUV waves were interacting 
with the streamer. Early  statistical  studies have found  that  the  coronal  magnetic  field  strength  falls  off  as  a
function of height, with an exponent of -1.5 within 0.02-9.0 solar radii (Dulk \& McLean, 1978; Lin et al., 2000). In the outer corona, at 
2.0 solar radii, the empirical relation of Dulk \& McLean (1978) gives the magnetic field value of $\approx$0.5 G, while Kwon et al. (2013) measures as 0.6 G. Our measured value, as diagnosed by first kink wave, estimates the magnetic field of 0.11$\pm$0.01 G at 2.0 solar radii, while it was 0.085$\pm$0.01 G as measured by second kink wave at the same height after one hour. Cho et al. (2007) observers 0.4 G magnetic field at a height of 2.2 solar radii, while our diagnosed values lie between 0.075 and 0.095 G. Chen et al. (2011) measures on average the magnetic field strength around 3.0 solar radii as 0.1 G, while our diagnosed values in the streamer at this height is $\approx$0.045 G. Keeping in view the uncertainty in measuring the electron density by Spherically Symmetric Polynomial Approximation (SSPA) method (Wang \& Davila, 2014), and also in the measurement of kink speed, it is obvious to have some mismatch with the typical magnetic field profiles of the outer corona. However, the diagnosed values of the magnetic field using principle of MHD seismology infer average magnetic field of streamer reasonably well with the value 0.1 Gauss at $>$ 2.0 solar radii. 

The power law for estimated magnetic field profile in the streamer are found to be $B_{s}$=$A$$\,r^{l}$, where $A$=0.54, 0.29 and $l$=-2.40, -1.84 in the difference of one hour time epoch. It is seen that during the evolution of second kink motion in the streamer, it increased in brightness (thus mass density; see Fig.~2), as well as in slight areal extent also, which may be associated with the decreased photospheric magnetic flux at the footpoint of the streamer (Poland et al., 1981). Therefore, it is most likely that the diagnosed magnetic field profile by second kink wave is reduced in value within the same streamer compared to the one diagnosed by first one. This kind of fine details about the temporal variation of magnetic field profiles even in the large-scale coronal structures (e.g., coronal rays, streamers, flux-ropes) can only be possible by its estimation using principle of MHD seismology. Therefore, MHD seismology of large-scale coronal structures, and thus the deduction of coronal magnetic fields has certain advantage over the classical methods of the magnetic field determination in the outer solar corona by assuming it to be isotropic (Dulk \& McLean, 1978; Lin et al., 2000). Additional achievement with coronal seismology is the precise measurement of streamer's magnetic field with uncertainty $<$10\%. By precise measurements of densities (e.g., with $<$5 \% error in the present case as estimated by SSPA inversion) as well as wave parameters (e.g., phase-speed with $<$10 \% error in the present measurements), we can constrain more accurate magnetic field values (e.g., with $<$10 \% error in present observations) of the outer coronal structures. Such accurate estimations of the magnetic field is very important as it could be utilized to understand more realistic coronal heating as well as solar wind models, and also in understanding the solar eruptions and their propagation in the outer corona. 

However, the utilized theory of MHD modes in the cylindrical fluxtube may not mimic exactly the morphology of the coronal waveguides (e.g., streamer here). Due to this underlying limitation of coronal seismology (CS), we need more realistic magnetic field
topology and density model in 3D MHD simulations of kink waves to invert the observational parameters in order to devise more accurate magnetic field. However, this work will be devoted to our future project, which will test the diagnosed magnetic field by using realistic model and observations, and will compare it with the one derived from basic method of coronal seismology.
\vspace{-0.7cm} 
\section{Acknowledgments}
AKS and BND acknowledge the RESPOND-ISRO (DOS/PAOGIA2015-16/130/602), and AKS acknowledges the SERB-DST project (YSS/2015/000621) grants. AKS and TS acknowledge respectively the Advanced Solar Computational \& Analyses Laboratory (ASCAL), and
infrastructural facilities at Department of Physics, IIT (BHU) to pursue this research. LO acknowledges support by NASA grant NNX16AF78G. Authors acknowledge the use of SSPA
technique developed by T.J. Wang and J.-M. Davila to measure the density in the outer corona. They also acknowledge the use of
STEREO EUVI and COR-1 observational data. We thank Dr. Tongjinag Wang for discussions and comments.

%
%



\end{document}